\documentclass[aps,prb,twocolumn,showpacs,10pt,nofootinbib]{revtex4-2}
\usepackage{graphicx}
\usepackage{amssymb}
\usepackage{amsmath,color,mathrsfs}
\usepackage{bbold}
\usepackage{comment}
\usepackage[dvipsnames]{xcolor}
\usepackage{bm} 
\pdfoutput=1
\usepackage{hyperref}
\hypersetup{urlcolor=blue, colorlinks=true, citecolor=blue, linkcolor=blue}

\usepackage{todonotes}
\usepackage{mdframed}
\graphicspath{{./fig/}}

\usepackage{overpic}[percent]

\usepackage{multirow}
\usepackage{booktabs}

\newcommand{\be}{\begin{equation}}
\newcommand{\ee}{\end{equation}}

\ifdefined\DeclareUnicodeCharacter
  \DeclareUnicodeCharacter{2219}{$\cdot$}
    \DeclareUnicodeCharacter{2212}{$-$}
    \DeclareUnicodeCharacter{22C5}{$\cdot$}
\fi

\begin{document}
\makeatother

\title{Quantification of the heavy-hole--light-hole mixing in two-dimensional hole gases}

\author{Peter Stano$^{1,2}$}
\email{peter.stano@riken.jp}
\author{Daniel Loss$^{1,3,4}$}

\affiliation{$^1$Center for Emergent Matter Science, RIKEN, 2-1 Hirosawa, Wako-shi, Saitama 351-0198, Japan}
\affiliation{$^2$Institute of Physics, Slovak Academy of Sciences, 845 11 Bratislava, Slovakia}
\affiliation{$^3$Department of Physics, University of Basel, Klingelbergstrasse 82, CH-4056 Basel, Switzerland}
\affiliation{$^4$RIKEN Center for Quantum Computing, 2-1 Hirosawa, Wako, Saitama, 351-0198 Japan}

\date{\today}

\newcommand{\dl}[1]{{\color{red}#1}}

\begin{abstract}

We theoretically investigate heavy-hole--light-hole mixing in two-dimensional hole gases (2DHG). We restrict our analysis to the zone center, appropriate for the low-density regime, which leads to a simple description, analytical results, and physical insights. We identify two different types of hole-Hamiltonian terms concerning mixing. The first type changes the direction of the pure spinors, without admixing light-hole components. It is efficient for Rabi driving the heavy-hole spin. The second type induces mixing and changes the eigenvalues of the $g$ tensor.  We analyze several measures that characterize the mixing quantitatively in Ge, Si, and GaAs, namely the $g$ factor, the light-hole weight in the wave function, the off-diagonal matrix elements in the Hamiltonian, and the strength of the induced spin-orbit interaction. 
We identify the canonical coordinate frame associated with a generic spin-3/2 Hamiltonian with time-reversal symmetry (TRS). In this coordinate frame, the mixing is quantified by a single parameter, the mixing angle $\vartheta$. We interpret it as the canonical (coordinate-frame and Hamiltonian-basis independent) measure of the heavy-hole--light-hole mixing. All the investigated mixing measures are simple functions of $\vartheta$.
As an illustration, we use our model to analyze heavy-hole spin qubit $g$ tensor, dephasing, relaxation, and Rabi frequencies, interpreting the arising effects as due to rotations of the canonical frame and changes of the mixing angle $\vartheta$.

\end{abstract}

\maketitle

\section{Introduction}

Holes in group-IV materials germanium (Ge) \cite{scappucci_germanium_2021} and silicon (Si) \cite{thompson_uniaxial-process-induced_2006} are currently among the prime candidates to host spin qubits \cite{loss_quantum_1998,fang_recent_2023}. They show excellent performance overall \cite{stano_review_2022} and even set the current state-of-the-art benchmarks for operation speed \cite{liu_ultrafast_2023}, operation fidelity \cite{lawrie_simultaneous_2023}, and qubit size \cite{lawrie_quantum_2020, borsoi_shared_2023}. Holes avoid the valley-degeneracy issue of the conduction band, couple weakly to nuclear spins \cite{prechtel_decoupling_2016}, and alleviate the use of micromagnets thanks to their strong spin-orbit interactions (SOIs) \cite{froning_ultrafast_2021}. 

The qualitative difference between electrons and holes as spin carriers originates in the band structure at the $\Gamma$ point of the zinc-blende and diamond semiconductors \cite{dresselhaus_cyclotron_1955, kane_energy_1956}. The higher (four-fold) degeneracy of the valence band makes holes anisotropic and much more responsive to electric and magnetic fields 
 \cite{bulaev_electric_2007,kloeffel_strong_2011,voisin_electrical_2016,crippa_electrical_2018,kloeffel_direct_2018, michal_longitudinal_2021}, as well as strain \cite{sun_physics_2007, terrazos_theory_2021}.
Judicious design of the wafer, device, and control fields can exploit this anisotropy to obtain well-performing \cite{hendrickx_single-hole_2020, hendrickx_four-qubit_2021} and tunable hole-spin qubits \cite{liles_electrical_2021, bosco_hole_2021,bosco_fully_2021,bosco_fully_2022, jirovec_dynamics_2022, hetenyi_anomalous_2022,michal_tunable_2023,geyer_anisotropic_2024}. There are many variants, stemming from qubits in clean Ge/SiGe heterostructures \cite{jirovec_singlet-triplet_2021, lodari_lightly_2022, kong_undoped_2023}, through high-speed and tunable ones in Ge hut \cite{watzinger_heavy-hole_2016,gao_sitecontrolled_2020, wang_ultrafast_2022} and selective-area growth \cite{ramanandan_coherent_2022} wires, Ge/Si core-shell wires \cite{higginbotham_hole_2014,zarassi_magnetic_2017,froning_strong_2021}, to industry-compatible Si finFET \cite{maurand_cmos_2016,kuhlmann_ambipolar_2018,camenzind_spin_2022} and MOS structures \cite{liles_spin_2018,ezzouch_dispersively_2021, piot_single_2022, yu_strong_2023}, to name just the main families.

\newcommand{\crystalAxis}[1]{{#1}}
\newcommand{\dotAxis}[1]{{#1^\prime}}
\newcommand{\deflectedAxis}[1]{{#1^{\prime\prime}}}
\newcommand{\someAxis}[1]{{\tilde{#1}}}

Despite the large variety of actual structures, the understanding of hole spin qubits starts with the following simple picture \cite{nenashev_wave_2003}. Consider a heterostructure quantum well (or an MOS epilayer surface) grown along the crystallographic [001] direction, denoted as the $\crystalAxis{z}$ axis.
The confinement splits the four-fold degenerate hole band \cite{martin_two-dimensional_1990} into two subbands, the heavy-hole (HH) and light-hole (LH) one, each with a two-fold spin (or Kramers) degeneracy. This splitting can be understood from the Luttinger model (given below), as the quantization energy of the quantum-well confinement. At the (Brillouin) `zone center', meaning putting the in-plane momenta to zero, $k_\crystalAxis{x}=0=k_\crystalAxis{y}$, the Luttinger Hamiltonian in the spin space reduces to $-J_\crystalAxis{z}^2$. Its eigenspinors are $(1,0,0,0)^T$ and $(0,0,0,1)^T$ with eigenvalue $-9/4$, and $(0,1,0,0)^T$ and $(0,0,1,0)^T$ with eigenvalue $-1/4$.\footnote{$\mathbf{J}$ is the (vector of operators of) total angular momentum of the Bloch wavefunction at the $\Gamma$ point, called in further `spin'. The basis to which the four-component spinors refer is the four Bloch wavefunctions with total angular momentum $3/2$, $1/2$, $-1/2$, and $-3/2$. See, for example, Tab. C1 on p. 208 in Ref.~\cite{winkler_spin-orbit_2003}.} The holes with the above spinors are sometimes referred to as  `pure heavy holes'.\footnote{The nomenclature is not firmly established. For example, the authors of Ref.~\cite{twardowski_variational_1987} (see their App.~A) call the superpositions $(1/\sqrt{2},0,0,\pm1/\sqrt{2})^T$ also `pure heavy holes'.} 

\makeatletter
\newcommand{\oset}[3][0ex]{%
  \mathrel{\mathop{#3}\limits^{
    \vbox to#1{\kern-2\ex@
    \hbox{$\scriptstyle#2$}\vss}}}}
\makeatother

\newcommand{\toHHa}[2]{\overset{#1}{#2}}%
\newcommand{\toHHb}{{hh \atop \to}} %
\newcommand{\toHHc}{\genfrac{}{}{-1pt}{}{hh}{\to}} %
\newcommand{\toHHd}{\oset[0.4ex]{hh}{\to}} %
\newcommand{\toHHe}[2]{\raisebox{-1pt}[0pt][0pt]{$\,\toHHa{#1}{#2}\,$}} %
\newcommand{\toHH}{\toHHe{\mathrm{hh}}{\to}}
\newcommand{\toTwoD}{\toHHe{\scriptscriptstyle \mathrm{2D}}{\to}} 

\newcommand{\mSOI}{m_\mathrm{soi}}%

The notable property of pure heavy holes is that the in-plane spin operators $J_\crystalAxis{x}$ and $J_\crystalAxis{y}$ do not couple them. Introducing $\mathbf{s} =\boldsymbol{\sigma}/2$, with $\boldsymbol{\sigma}$ the Pauli matrices, as the pseudo-spin 1/2 operators in the two-dimensional subspace spanning the heavy-hole spinors, one has the projection rule
\be
\label{eq:projectionRules0}
J_\crystalAxis{x} \toHH 0, \quad J_\crystalAxis{y} \toHH 0,  \quad  J_\crystalAxis{z} \toHH 3 s_\crystalAxis{z}.
\ee
This rule is the basis for anisotropy of various spin-related properties. For example, the bulk Zeeman interaction $2 \mu_B \kappa\, \mathbf{J} \cdot \mathbf{B}$ projects to $6 \mu_B \kappa s_\crystalAxis{z} B_\crystalAxis{z}$, so that pure heavy holes have strongly anisotropic $g$ factors ($6 \kappa$ out of the plane, zero in the plane). Essentially the same argument gives similarly anisotropic hyperfine interaction, $s_\crystalAxis{z} I_\crystalAxis{z}$, between the hole and nuclear spin $\mathbf{I}$. Or, pure heavy holes should have cubic-in-momenta spin-orbit interaction (SOI)~\cite{bulaev_spin_2005-1,bulaev_electric_2007}, since the bulk linear-in-momenta interactions (such as Rashba) that contain $J_\crystalAxis{x}$ and $J_\crystalAxis{y}$ operators are projected to zero.

Going beyond the above simplistic model, the spinors of heavy holes deviate from the pure ones. This is called `heavy-hole--light-hole mixing' and can arise in various ways. On the one hand, there is mixing even at the zone center (that is, still with $k_\crystalAxis{x}=0=k_\crystalAxis{y}$) if the growth direction is deflected away from [001]  \cite{winkler_highly_2000}, if there is strain \cite{nenashev_wave_2003,thompson_uniaxial-process-induced_2006},
 or from heterostructure interfaces \cite{ivchenko_heavy-light_1996}. On the other hand, even without strain or low-symmetry confinement, mixing arises upon going away from the zone center, once the in-plane momenta become appreciable. Heavy holes might still be close to being pure in lateral quantum dots if $k_\crystalAxis{x} \sim k_\crystalAxis{y} \ll k_\crystalAxis{z}$, while the pure-heavy-hole picture breaks down completely in quantum wires \cite{kloeffel_direct_2018} for which the momentum hierarchy is $k_\crystalAxis{x} \ll k_\crystalAxis{y}\sim k_\crystalAxis{z}$. The spinor structure of holes then depends sensitively on the wire cross section \cite{adelsberger_hole-spin_2022,bosco_hole_2022}. 

In a typical gated spin-qubit device, all these mixing sources are present and result in holes with complex and versatile spin character. In this article, we analyze the heavy-hole--light-hole mixing\footnote{For conciseness, we will often say `mixing', dropping the `heavy-hole--light-hole' quantifier. Upon mixing, the hole spinors cease to be `pure'.} arising from strain or from the two-dimensional hole gas (2DHG) normal being deflected from the [001] direction. We use a simple model appropriate for $k_\crystalAxis{z} \gg k_\crystalAxis{x}, k_\crystalAxis{y}$ (we call it `zone-center model', see below), as a minimal extension of the above pure-heavy-hole model applicable to 2DHG (at low density) and nanostructures based in it, such as lateral quantum dots in the single-hole regime.

Our contribution is in investigating the \emph{quantitative degree} of some mixing measures that are motivated by experiments and theory. Namely, we analyze 1) the heavy-hole $g$ factor, 2) pure light-hole admixture in the spinor wave function, 3) the sum of the squares of the off-diagonal elements of the Hamiltonian matrix, and 4) the degree of breaking the projection rule $J_\crystalAxis{x}, J_\crystalAxis{y} \toHH 0$ resulting in appearance of the spin-orbit interactions. Interestingly, we find that a simple additional requirement of coordinate-frame independence results in measures 1), 2), and 3) to pick the same coordinate frame. We interpret it as the canonical coordinate frame associated with a spin-3/2 Hamiltonian with time-reversal symmetry (TRS). It defines a single heavy-hole--light-hole mixing parameter, which we express as an angle $\vartheta \in [0,\pi/2]$.\footnote{The domain can be restricted to $\vartheta \in [0,\pi/6]$, see below (page \pageref{page:varThetaDomain}).} All four investigated measures are simple functions of this mixing angle $\vartheta$.

The proposal of $\vartheta$ as the canonical mixing parameter is our main result. Two more points are worth adding here. First, concerning mixing, not all terms in the Hamiltonian are equally efficient: the terms $J_\crystalAxis{z} J_\pm$ give much less mixing than terms $J_\pm^2$. The reason is that the dominant effect of the former ones is only a rotation of the canonical frame. In this frame, the spinors remain pure and the projection rule $J_\crystalAxis{x}, J_\crystalAxis{y} \to 0$ still holds. 
Second, the mixing is more efficient in inducing the SOI, quantified by $m_4$, which is linear in $\vartheta$, than for the effects quantified by $m_1$, $m_2$, and $m_3$, which are quadratic in $\vartheta$.  The difference is pronounced at small mixing: heavy holes which are as much as 95-98\% `pure', if considering the pure heavy-hole content or the $g$ factor difference to 3/2, can have effective Rashba SOI,
\be
\nonumber
\alpha_r^\mathrm{\scriptscriptstyle 3D} E_\crystalAxis{z} (J_\crystalAxis{x} k_\crystalAxis{y} - J_\crystalAxis{y} k_\crystalAxis{x}) \,\,\, \toHH \,\,\, \alpha_r^\mathrm{\scriptscriptstyle 2D} E_\crystalAxis{z} (s_\crystalAxis{x} k_\crystalAxis{y} - s_\crystalAxis{y} k_\crystalAxis{x}),
\ee
with the strength comparable to the bulk SOI strength, $\alpha_r^\mathrm{\scriptscriptstyle 2D} = O( \alpha_r^\mathrm{\scriptscriptstyle 3D})$.\footnote{
The difference can be seen from Fig.~\ref{fig:varThetaFigure}.
} This property holds for the mixing of any source. Converting to numerical values for strain as an example, in a quantum dot where the heavy-hole--light-hole splitting is dominated by an in-plane compressive strain, such as the Ge/Si$_{0.2}$Ge$_{0.8}$ quantum well, the off-diagonal $\epsilon_{xy}$ strain components an order of magnitude smaller than the built-in strain $\epsilon_{xx} \approx -0.6\%$ \cite{abadillo-uriel_hole-spin_2023,sammak_shallow_2019} will also result in a heavy-hole Rashba SOI  strength comparable to its value in bulk.

We now proceed to derivations and quantitative details. While our analysis applies to holes in generic crystals with zinc-blende or diamond structure, in the main text we present results for silicon and germanium. For completeness, in some figures we include GaAs.

\section{The pure heavy-hole limit}

We start with the Luttinger Hamiltonian describing the bulk top valence band, the spin-3/2 hole band \cite{luttinger_quantum_1956},
\begin{equation}
	\begin{split}
		H_{L}  = & \,  \left(\gamma_1 +\frac{5}{2}\gamma_2\right) \frac{\hbar^2 k^2}{2m_0}
		 - \gamma_2 \frac{\hbar^2}{m_0} \left( k_\crystalAxis{x}^2 J_\crystalAxis{x}^2+\mathrm{c.p.}\right) \\
		&-\gamma_3 \frac{\hbar^2}{2m_0}( \{k_\crystalAxis{x},k_\crystalAxis{y}\}\{J_\crystalAxis{x},J_\crystalAxis{y}\} +\mathrm{c.p.}).
	\end{split}
\label{eq:Luttinger}
\end{equation}
Here, $\gamma_i$ are the Luttinger parameters, $k_\alpha$ are Cartesian components of the momentum operator  in crystallographic coordinates $\alpha \in \{ \crystalAxis{x}=[100],\crystalAxis{y}=[010], \crystalAxis{z}=[001] \}$, $J_\alpha$ are spin-3/2 operator Cartesian components, $\{A,B\} = AB+BA$ is the anticommutator, and c.p.~stands for cyclic permutations (of Cartesian components). We take the hole band facing up, so that higher excited states have higher energies.

We now consider the effects of confinement $V(\crystalAxis{z})$, say a quantum well, along the growth direction $\crystalAxis{z}$. Aiming at a simple model, we assume that the arising quantization of the momentum along $\crystalAxis{z}$ dominates the in-plane quantization, $k_\crystalAxis{x}, k_\crystalAxis{y} \ll k_\crystalAxis{z}$.\footnote{We momentarily treat $k_i$ as c-numbers, even though they are operators in Eq.~\eqref{eq:Luttinger}. A more precise statement (applicable for 2DHG, where $k_x$ and $ k_y$ can be treated as c-numbers), would be $k_x l_z, k_y l_z \ll 1$, with $l_z$ the confinement length defined by $V(z)$.}
Accordingly, we put $k_\crystalAxis{x}=0=k_\crystalAxis{y}$ in Eq.~\eqref{eq:Luttinger},\footnote{
We note that by this step we remove also the so-called `direct Rashba SOI'~\cite{kloeffel_strong_2011} from our model. This type of SOI  originates from the mixed terms of $H_L$, such as  $\{k_\crystalAxis{x},k_\crystalAxis{z}\}\{J_\crystalAxis{x},J_\crystalAxis{z}\}$, %
and from an electric field that breaks inversion symmetry~\cite{kloeffel_strong_2011,kloeffel_direct_2018,bosco_hole_2021}.} and get
\begin{equation}
\label{eq:unperturbedSpinorH0}
	\begin{split}
		H_0  &= H_{L}(\mathrm{zone\, center})  + V(\crystalAxis{z})\\
		&=  \, \frac{\hbar^2  k_\crystalAxis{z}^2}{2m_0}\left( \gamma_1 +\frac{5}{2}\gamma_2 - 2\gamma_2 J_\crystalAxis{z}^2 \right) +V(\crystalAxis{z}).
	\end{split}
\end{equation}
In further, we will refer to the approximation $\mathbf{k}_{\|}=(k_x,k_y)=0$ as the `zone center model', for simplicity.
One advantage of restricting the analysis to the zone center is that the spinor structure does not depend on the confinement potential and the associated orbital energies. We can thus  focus on the spin operator in the kinetic energy, which we denote by $H_S$, and ignore the orbital degrees of freedom. In Eq.~\eqref{eq:unperturbedSpinorH0}, the operator is 
\be
\label{eq:HS}
H_S = \gamma_1 +\frac{5}{2}\gamma_2 - 2\gamma_2 J_\crystalAxis{z}^2,
\ee
and its eigenstates are
\begin{subequations}
\label{eq:pureHoleSpinors}
\begin{align}
&\left. \begin{tabular}{c} $|\Psi_{+3/2}^\mathrm{pure} \rangle = (1,0,0,0)^T$ \\ $|\Psi_{-3/2}^\mathrm{pure}\rangle  = (0,0,0,1)^T$
	\end{tabular} \right\} \mathrm{pure\, heavy\, holes},\\
	&\qquad E_{\pm 3/2} = \gamma_1+\frac{5}{2}\gamma_2 -\frac{9}{2}\gamma_2,\\\nonumber\\
&\left. \begin{tabular}{c} $|\Psi_{+1/2}^\mathrm{pure} \rangle = (0,1,0,0)^T$ \\ $|\Psi_{-1/2}^\mathrm{pure} \rangle = (0,0,1,0)^T$
	\end{tabular} \right\}  \mathrm{pure\, light\, holes},\\
	&\qquad E_{\pm 1/2} = \gamma_1+\frac{5}{2}\gamma_2 -\frac{1}{2}\gamma_2.
\end{align}
We will call the difference of the energies,
\be
\label{eq:Delta}
\Delta = E_{\pm1/2}-E_{\pm 3/2} = 4\gamma_2,
\ee
\end{subequations}
the heavy-hole--light-hole splitting. Even though this is not precise\footnote{The energy splitting of the eigenstates of Eq.~\eqref{eq:unperturbedSpinorH0} is not given as $\Delta$ times an orbital energy scale. It becomes so only upon neglecting the influence of the last term in the bracket in Eq.~\eqref{eq:unperturbedSpinorH0} on the particle mass.}, $\Delta$ is the important parameter for the spin structure of holes, our focus.

Going beyond the model in Eq.~\eqref{eq:unperturbedSpinorH0}, the hole spinors will not be equal to the pure hole spinors in Eq.~\eqref{eq:pureHoleSpinors}. We denote these general spinors by $\Psi_{\pm 3/2}, \Psi_{\pm 1/2}$. We define the vector of heavy-hole pseudo-spin-1/2 operators (Pauli matrices) $\boldsymbol{\sigma}=\{\sigma_\crystalAxis{x}, \sigma_\crystalAxis{y}, \sigma_\crystalAxis{z}\}$, and alternatively $\mathbf{s}=\boldsymbol{\sigma}/2$, to act in the two-dimensional subspace spanned by $\{ \Psi_{+3/2}, \Psi_{-3/2}$ \} with the pseudo-spin `up' corresponding to $\Psi_{+3/2}$ and `down' to $\Psi_{-3/2}$.

\section{The measures of mixing}

With a growth direction other than a high-symmetry axis, or upon including strain or finite in-plane momenta, the spinor-defining operator becomes more complicated than $J_\crystalAxis{z}^2$ and its eigenstates differ from the pure hole spinors in Eq.~\eqref{eq:pureHoleSpinors}.\footnote{Using the Broido-Sham transformation \cite{broido_effective_1985,lee_effects_1988,fishman_hole_1995}, one can obtain analytical eigenspinors for all Hamiltonians considered in this paper. However, we do not find the resulting formulas helpful for physical insight and do not pursue this direction.} We are interested in having a measure for this difference. 

\subsection{Some plausible measures of mixing}

Let us consider the following definition\footnote{\label{fnt:TRS}Throughout this paper, we assume that the heavy-hole subspace is identified using a Hamiltonian that has time-reversal symmetry (TRS). In this case, an alternative definition $1 + (2/3) \textrm{min} \cdots$ is equivalent to the expression $1-(2/3)\textrm{max} \cdots$ given in Eq.~\eqref{eq:measure1}.} 
\be
\label{eq:measure1}
m_1(\Psi_{\pm 3/2}) =1- \frac{2}{3}\max_{\mathbf{n}, \alpha,\beta} \langle \Psi_\mathrm{hh}^{\alpha\beta} | \mathbf{J} \cdot \mathbf{n} | \Psi_\mathrm{hh}^{\alpha\beta} \rangle,
\ee 
where $\mathbf{n}$ is a real unit vector, $\alpha$ and $\beta$ fulfilling $|\alpha|^2 + |\beta|^2=1$ are complex numbers, and 
\begin{equation}
\label{eq:Psihh}
|\Psi_\mathrm{hh}^{\alpha\beta}\rangle = \alpha | \Psi_{+3/2}\rangle + \beta | \Psi_{-3/2} \rangle,
\end{equation} 
is a normalized spinor within the heavy-hole subspace.\footnote{The maximization over coefficients $\alpha$ and $\beta$ can be replaced by evaluating the eigenvalues of the $2 \times 2$ -matrix of the projection of $\mathbf{J} \cdot \mathbf{n}$ to the subspace spanned by $\Psi_{\pm 3/2}$. Due to TRS, the eigenvalues come in pairs $\pm \epsilon$ and the coefficients $\{\alpha ,\beta\}$ maximizing the expectation value correspond to the eigenvector with the eigenvalue $+\epsilon$.} The mixing is measured as the minimal possible decrease of the expectation value of the spin from its maximum 3/2, if one can choose any suitable direction for the axis along which the spin is measured and any suitable spinor from the heavy-hole subspace. In this way, the quantity $1-m_1$ gives the maximal $g$ factor [due to the linear Zeeman term, Eq.~\eqref{eq:Hz3D}] that can be measured in a state from the heavy-hole subspace. Such a quantity can be accessed experimentally by measuring the $g$ factor as a function of the orientation of the magnetic field of a fixed modulus. Thus, the choice of $m_1$ is motivated by current experimental techniques.

We next examine the following option
\be
\label{eq:measure2}
m_2(\Psi_{\pm 3/2}) = 1-\frac{1}{2}\max_{\mathbf{n}} \mathrm{Tr}[ \rho_{\pm 3/2} \rho_{\pm 3/2}^\mathrm{pure}(\mathbf{n})],
\ee 
where $\rho_{\pm 3/2} = |\Psi_{+3/2}\rangle \langle \Psi_{+3/2}| + |\Psi_{-3/2}\rangle \langle \Psi_{-3/2}|$ is the projector to the heavy-hole subspace. The projector $\rho_{\pm 3/2}^\mathrm{pure}$ is analogous, but using pure heavy-hole spinors, Eq.~\eqref{eq:pureHoleSpinors}, defined by choosing their `$z$' axis along a unit vector $\mathbf{n}$. The value of $m_2$ is obtained upon finding the direction $\mathbf{n}$ that defines a pure basis with the maximal overlap with the exact spinors. This choice suits analytics and numerics where the spinor wave functions are available. 

As the third example, we consider a measure based on the Hamiltonian, rather than its explicit eigenstates. We define the  $4\times 4$  matrix $h^\mathbf{n}$ as the Hamiltonian evaluated in a basis of pure spinors that are defined by an arbitrary vector $\mathbf{n}$ defining the `$z$' axis. The sum
\be
\label{eq:measure22}
m_3 (H) = \frac{1}{\Delta^{2}}\min_{\mathbf{n}} \sum_{i\neq j} |h^{\mathbf{n}}_{ij}|^2
\ee 
evaluates the minimal sum of absolute squares of the off-diagonal elements of the Hamiltonian. We normalize it by dividing it by the available energy scale $\Delta$.

\begin{center} \textbf{Canonical reference frame} \end{center}

The examples suggest that one can come up with many variants of a `mixing measure', with preference depending on the application. Nevertheless, there are certain natural requirements. First, if the pure heavy-hole spinors are the system eigenstates, the mixing should be zero, $m=0$. Second, the measure should be independent of the coordinate system in which the spinors or the Hamiltonian are evaluated. Third, the measure should be invariant under unitary rotations of the basis within the heavy-hole spinor subspace. All three measures, $m_i, i=1,2,3$, fulfill these minimal requirements.

In addition, and perhaps surprisingly, all  measures $m_i$ select the same vector $\mathbf{n}$. We delegate the details and proofs to App.~\ref{app:proofs} and give the summary here. In the coordinate frame with the `$z$'  axis along the vector $\mathbf{n}$, the spinor-defining Hamiltonian $H_S$ evaluated in the basis composed of pure spinors takes the following form (up to an additive constant and an overall multiplicative factor)
\be
\label{eq:H44_simple}
\begin{tabular}{c|c}
\begin{tabular}{cccc}
$\scriptstyle{+3/2}$ \\$\scriptstyle{+1/2}$\\$\scriptstyle{-1/2}$\\$\scriptstyle{-3/2}$
\end{tabular} 
&
$\left(
\begin{tabular}{cccc}
$-1$ & $0$ & $\alpha$ & $0$\\
$0$ & $1$ & $0$ & $\alpha$\\
$\alpha^*$ & $0$ & $1$ & $0$\\
$0$ & $\alpha^*$ & $0$ & $-1$
\end{tabular} 
\right).$
\end{tabular}
\ee
There is a single non-trivial matrix element $\alpha$ off the diagonal, which can be parametrized by two real numbers. We will use two angles, $\vartheta \in [0,\pi/2]$ and $\varphi \in[0,2\pi]$, introducing
\be
\label{eq:alpha}
\alpha = e^{-i \varphi} \tan\vartheta.
\ee
The  measures $m_i$ can be expressed as simple functions of the parameter $\vartheta$,
\begin{subequations}
\begin{align}
\label{eq:m1Result}
m_1 &= \sin^2 (\vartheta/2), &\vartheta \leq \pi/2,&\\
\label{eq:m2Result}
m_2 &= 2\sin^2 (\vartheta/2),&\vartheta \leq \pi/3,\\
\label{eq:m4Result}
m_3 &= \tan^2 \vartheta,  &\vartheta \leq \pi/6.
\end{align}
\end{subequations}

We interpret these findings as follows. For every spin-3/2 Hamiltonian with time-reversal symmetry, there is a canonical coordinate frame, defined by the vector $\mathbf{n}$ as the frame `$z$' axis, and the inherent heavy-hole--light-hole mixing angle $\vartheta$. The quantities $\mathbf{n}$ and $\vartheta$ are well-defined in the sense that they are independent of the original coordinate frame, Hamiltonian basis, or the chosen measure $m_i$  to obtain them. In the canonical coordinate frame, the Hamiltonian is (in a certain sense) the simplest possible, the $g$ factor of the heavy-hole spinor is maximal possible with the magnetic field along the frame preferred axis, and the exact spinor has maximal possible overlap with a pure heavy-hole spinor, the latter defined along the preferred axis. 

\begin{center} \textbf{Canonical mixing measure} \end{center}

The above results also suggest to consider $\alpha$ or, more specifically, the angle $\vartheta$ as the canonical mixing measure. We call $\vartheta$ the mixing angle. Indeed, all considered measures $m_i$, including $\mSOI$ given below, are simple monotonic functions of the mixing angle. The mixing angle refers to the canonical frame and thus fulfills the invariance requirements discussed  above. 

The results of the investigations of the measures $m_i$ imply the following clarification of the definition range of the mixing angle. In Eq.~\eqref{eq:H44_simple}, the parameter $\alpha$ can be arbitrary, giving a valid time-reversal-symmetric spin-3/2 Hamiltonian. Correspondingly, we defined the mixing angle on the interval $\vartheta \in [0,\pi/2]$ in Eq.~\eqref{eq:alpha}. However, while the canonical reference frame is uniquely defined by that Hamiltonian, the value of parameter $\alpha$ can be changed by renaming the axes of the canonical frame. With such a renaming, one can map a problem with $\vartheta > \pi/3$ to a problem with $\vartheta <\pi/3$, with the explicit mapping $\vartheta \to 2\pi/3-\vartheta$, see Eq.~\eqref{eq:alphaMapping}. This availability can be noticed by studying $m_2$, which displays a kink at $\vartheta = \pi/3$. Similarly, $m_3$ reveals a kink at $\vartheta=\pi/6$, reflecting yet another reordering of the canonical frame axes. With this third choice, the diagonal part of the Hamiltonian becomes $+J_\deflectedAxis{z}^2$ and describes a system with light holes as the ground state. Converting Eq.~\eqref{eq:H44_simple} to such a form implies a transformation $\vartheta \to \pi/3-\vartheta$. We thus conclude that a TR-symmetric spin-3/2 Hamiltonian, which can always be brought to the form given in Eq.~\eqref{eq:H44_simple}, can be  interpreted as a system where the heavy-hole ground state is perturbed by the heavy-hole--light-hole mixing only for $|\alpha| \leq 1/\sqrt{3}$, corresponding to $\vartheta \leq \pi/6$. \label{page:varThetaDomain}

\begin{figure}
\begin{center}
\includegraphics[width=0.9\linewidth]{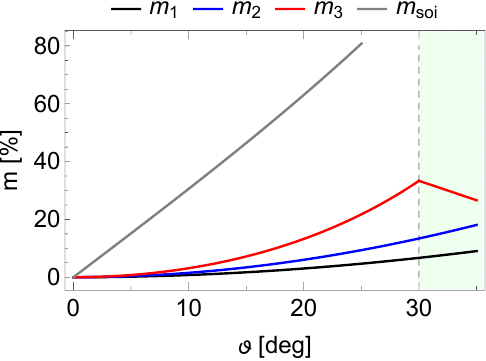}\\
\end{center}
\caption{
Plot of the four measures $m_1$, $m_2$, $m_3$, $\mSOI$, quantifying the mixing of heavy holes with light holes, as function of the mixing angle $\vartheta$. The range $0 \leq \vartheta \leq \pi/6$ corresponds to a system with the heavy-hole ground state perturbed by the heavy-hole--light-hole mixing. For larger $\vartheta$ (shaded area of the plot), a more fitting description of the system would be to take a light-hole subspace as the ground state perturbed by mixing (we do not pursue such analysis in this article). The measure $m_3$ features a kink at the transition. There is a similar kink for $m_2$ at $\vartheta=\pi/3$ (not shown). 
\label{fig:varThetaFigure}
}
\end{figure}

We plot the four investigated measures in Fig.~\ref{fig:varThetaFigure} for the mixing-angle range corresponding to such weakly-perturbed heavy-hole systems.

\subsection{Mixing measure related to SOI}

We now introduce one more mixing measure of a different type, quantified and motivated as follows. We assume that there is a linear-in-$J$ and linear-in-$k$ SOI  in the bulk, taken as Rashba SOI for concreteness,
\be
\label{eq:bulkRashba}
		H_r^\mathrm{\scriptscriptstyle 3D}  =  \alpha_r^\mathrm{\scriptscriptstyle 3D}\, 	\mathbf{E} \cdot (\mathbf{J} \times \mathbf{k}).
\ee
Assuming that the electric field $\mathbf{E}$ is along the growth direction $\crystalAxis{z}$ and using Eq.~\eqref{eq:pureHoleSpinors}, one finds that for pure heavy holes this interaction is projected to zero within the subspace defined by pure heavy-hole spinors. However, with mixing, SOIs appear within the heavy-hole subspace \cite{rashba_spin-orbital_1988, habib_negative_2004}.
 Though not exclusively (see below), typically they take the form analogous to the bulk form,\footnote{On going from Eq.~\eqref{eq:bulkRashba} to Eq.~\eqref{eq:2DHGRashba}, our simple procedure includes only the homogeneous electric field. Rather than a quantitative theory of SOI in 2DHG, our focus is to size up the effects of mixing. The quantitative theory of SOI requires a more elaborate approach, due to the appearance of additional contributions from the quantum well interfaces. While the bands discontinuities arising at the interfaces can be viewed as electric fields, they cannot be simply added to $\mathbf{E}$ since they are band dependent. This is a well-known issue and the nature of $\mathbf{E}$ in Eq.~\eqref{eq:2DHGRashba} has been discussed at length elsewhere, see, for example, p.~99 in Ref.~\onlinecite{fabian_semiconductor_2007}.}
\be
\label{eq:2DHGRashba}
		H_r^\mathrm{\scriptscriptstyle 2D}  =  \alpha_r^\mathrm{\scriptscriptstyle 2D}\, 	\mathbf{E} \cdot (\mathbf{s} \times \mathbf{k}).
\ee
Our third measure is the ratio of the two constants,
\be
\label{eq:measure3}
\mSOI(\Psi_{\pm 3/2}) = \frac{\alpha_r^\mathrm{\scriptscriptstyle 2D}}{\alpha_r^\mathrm{\scriptscriptstyle 3D}}.
\ee 
We do not define this measure through a minimization procedure. Instead, we evaluate it in the canonical frame and find [see Eq.~\eqref{eq:m3Exact}] that it also is a simple function of the mixing angle,
\be
\label{eq:mSOIResult}
\mSOI = \sqrt{3} \tan \vartheta.
\ee
Next, we evaluate the mixing in specific scenarios.

\section{Mixing due to a deflected 2DHG normal}

\label{sec:deflection}

\newcommand{\growthDirection}{{2DHG normal}}

We now consider a 2DHG with normal deflected from the high-symmetry direction [001]. First of all, it describes a 2DHG implemented in heterostructures with nominal growth direction different from [001]. However, we are additionally motivated by experiments where the holes are trapped by confinement that is appreciably modulated by gates and interfaces. Unlike in heterostructures buried deep below the surface with the gate-stack layers, in this scenario, the nominal growth direction of the crystal can be different from the normal of the effective plane confining the 2DHG. The examples include MOS structures with gates (and impurities) very close to the 2DHG and finFET structures where the holes are pushed against a surface that is not planar \cite{piot_single_2022, camenzind_spin_2022}. To stress this interpretation, we will use `(effective) \growthDirection{}' instead of `growth direction'. The name refers to the effective 2DHG confinement at the location of a quantum dot, rather than the nominal growth direction of the bulk crystal.

With the \growthDirection{} deflected from [001], the mixing arises even at the band center \cite{winkler_rashba_2000}. We first consider the \growthDirection{} along a general direction [klh]. We define rotated coordinates $\dotAxis{x},\dotAxis{y},\dotAxis{z}$ by 
\begin{subequations}
\label{eq:coordinateRotation}
\be
\left( \begin{tabular}{c} $\crystalAxis{x}$ \\$\crystalAxis{y}$\\$\crystalAxis{z}$ \end{tabular} \right) = R[\phi, \theta,\phi^\prime] \left( \begin{tabular}{c} $\dotAxis{x}$ \\$\dotAxis{y}$\\$\dotAxis{z}$ \end{tabular} \right),
\ee
where the rotation is parametrized by three Euler angles
\be
\label{eq:rotationMatrixR}
R[\phi, \theta,\phi^\prime] =  \left( \begin{tabular}{ccc} 
$c_\theta c_\phi c_{\phi^\prime} -s_\phi s_{\phi^\prime} $ & $-c_\theta c_\phi s_{\phi^\prime} -s_\phi c_{\phi^\prime} $& $c_\phi s_\theta $\\ $c_\theta s_\phi c_{\phi^\prime} +c_\phi s_{\phi^\prime} $ & $-c_\theta s_\phi s_{\phi^\prime} +c_\phi c_{\phi^\prime} $& $s_\phi s_\theta $\\ 
$-s_\theta c_{\phi^\prime} $ & $s_\theta s_{\phi^\prime}$& $c_\theta $\\ \end{tabular} \right),
\ee
\end{subequations}
with $s_u \equiv \sin u$ and $c_u \equiv \cos u$. In this parameterization, the \growthDirection{} (the $\dotAxis{z}$ axis) is along the unit vector with nonprimed (crystallographic) coordinates given by the last column of the Euler rotation matrix $R$. 

\subsection{Growth directions [llh] and [0lh]}

\label{sec:scenarios}

\begin{figure}
  \begin{overpic}[scale=0.95]{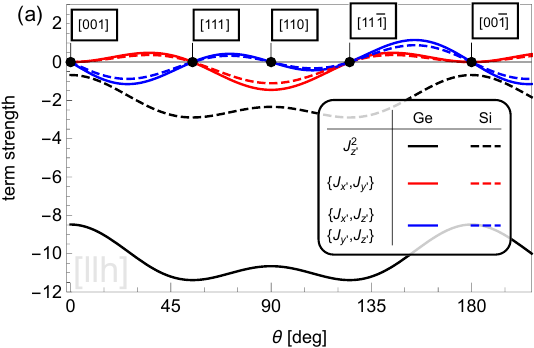}
     \put(22,13){\includegraphics[scale=0.4]{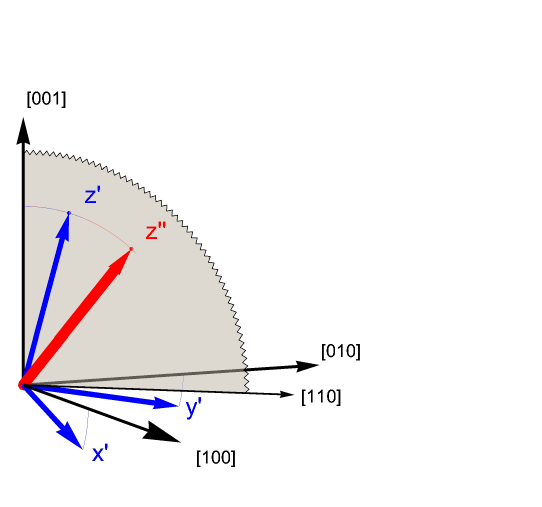}}  
  \end{overpic}\\
  \vspace{0.5cm}
  \begin{overpic}[scale=0.95]{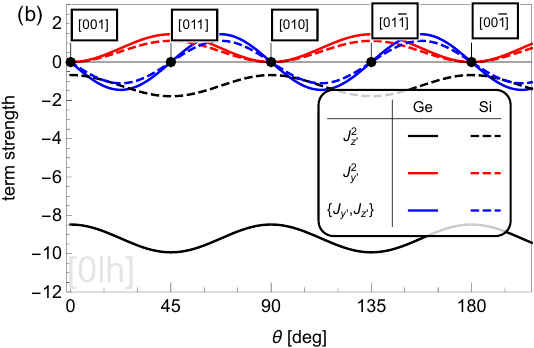}
     \put(18,14){\includegraphics[scale=0.4]{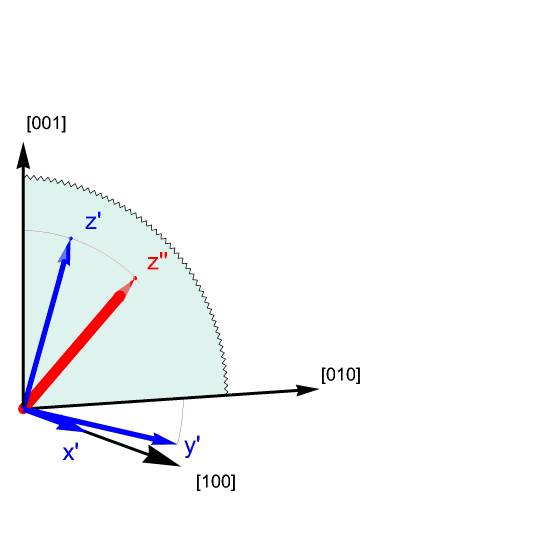}}  
  \end{overpic}
\caption{
\textbf{The zone-center Hamiltonian}. The amplitudes of the spin-dependent terms in (a) Eq.~\eqref{eq:H0-firstScenario} and (b) Eq.~\eqref{eq:H0-secondScenario} as a function of the angle $\theta$ defining the \growthDirection{} direction. Several high-symmetry directions are denoted by points with boxes. Solid lines are for the parameters of Ge, dashed for Si. For the direction [llh], the strength of the terms $\{ J_\dotAxis{x}, J_\dotAxis{z}\}$ and $\{ J_\dotAxis{y}, J_\dotAxis{z}\}$ are the same; only one is plotted. The insets show the coordinate axes: the crystallographic axes are in black, the rotated coordinate system with $z^\prime$ the \growthDirection{} is in blue, and the canonical coordinate frame $z^{\prime\prime}$ axis is in red. The orientations of $z^\prime$ and $z^{\prime\prime}$ are to scale for the parameters of Si for $\theta \approx 16^\circ$.
\label{fig:spinOperatorParts}
}
\end{figure}

In rotated coordinates, the zone-center spinor-defining Hamiltonian is still bilinear in the spin operators,
\begin{subequations}
\label{eq:H0-delfectedGrowthDirection}
\be \label{eq:H0-delfectedGrowthDirection-coefficientsC} 
H_{S}^{[klh]} =
\sum_{i,j\in {\dotAxis{x},\dotAxis{y},\dotAxis{z}}} \frac{c_{ij}}{2} \{J_i,J_j\},
\ee
with the coefficients $c_{ij}$ being functions of the Euler angles and Luttinger parameters, and the factor 2 is introduced for later convenience.
Since the most general case leads to unwieldy expressions and figures, we exemplify it by two specific scenarios with a single rotation parameter. First, we consider rotations around $\mathbf{m}=[110]$ by an arbitrary angle $\theta$. It corresponds to $\phi=\pi/4$ and $\phi^\prime=-\pi/4$ in Eq.~\eqref{eq:coordinateRotation}. The choice covers the most typical cases, [001] for $\theta=0$, [111] for $\theta=\arctan(\surd{2})$, and [110] for $\theta=\pi/2$, %
as well as the generic [llh] direction investigated at length both experimentally and theoretically.\footnote{See, for example, Ref.~\cite{winkler_highly_2000} and the references in the introductions of Refs.~\cite{fishman_hole_1995,winkler_anisotropic_1996}.}
The Hamiltonian is (a spin-independent constant is omitted)
\be \label{eq:H0-firstScenario} 
\begin{split}
H_{S}^{[llh]}&=
 \left(-2\bar{\gamma} -\gamma_\delta\frac{5-12\cos2\theta-9\cos4\theta}{8}\right) J_\dotAxis{z}^2 \\
&- \gamma_\delta \sin\theta \frac{5\cos\theta+3\cos3\theta}{2\sqrt{2}} \{ J_\dotAxis{x}+J_\dotAxis{y},J_\dotAxis{z}\}\\
&+\gamma_\delta \sin^2\theta (3\cos2\theta + 1) \{ J_\dotAxis{x},J_\dotAxis{y}\},
\end{split}
\ee
where we introduce $\bar{\gamma}=(\gamma_3+\gamma_2)/2$ and $\gamma_\delta=(\gamma_3-\gamma_2)/2$.\footnote{Our definition of $\bar{\gamma}$ is in line with Refs.~\cite{lipari_angular_1970,broido_effective_1985,foreman_analytic_1994,fishman_hole_1995,budkin_spin_2022}. Unfortunately, there is no agreed-on convention for $\gamma_3-\gamma_2$. For example, Ref.~\cite{fishman_hole_1995} uses $\delta=(\gamma_3-\gamma_2)/\gamma_1$, Refs.~\cite{broido_effective_1985,foreman_analytic_1994} use $\mu=(\gamma_3-\gamma_2)/2$,  Ref.~\cite{fishman_hole_1995} uses $\gamma_\delta=(\gamma_3-\gamma_2)/2$, Ref.~\cite{winkler_highly_2000} uses $\gamma_\delta=(\gamma_3-\gamma_2)$, and Ref.~\cite{budkin_spin_2022} uses $\gamma_3-\gamma_2$ without giving it a name.}
The second scenario is a rotation around $\mathbf{m} = [\overline{1}00]$. This corresponds to Eq.~\eqref{eq:coordinateRotation} with $\phi=\pi/2$ and $\phi^\prime=-\pi/2$, and includes [001] for $\theta=0$ and [011] for $\theta=\pi/4$, as well as a generic direction [0lh] investigated, for example, in Ref.~\cite{budkin_spin_2022}. The Hamiltonian is (constant omitted)
\be \label{eq:H0-secondScenario} 
\begin{split}
H_{S}^{[0lh]}
&=  \big(-2\bar{\gamma} +\gamma_\delta(1+\cos4\theta)\big) J_\dotAxis{z}^2 \\
&- 2\gamma_\delta \sin\theta (\cos\theta+\cos3\theta) \{J_\dotAxis{y},J_\dotAxis{z}\}\\
&+2\gamma_\delta \sin^2 (2\theta) J_\dotAxis{y}^2.
\end{split}
\ee
\end{subequations}
The amplitudes of the terms from Eqs.~\eqref{eq:H0-firstScenario} and \eqref{eq:H0-secondScenario} are plotted in Figs.~\ref{fig:spinOperatorParts}(a) and (b), respectively. In both scenarios, $J_\dotAxis{z}^2$ dominates for any direction of the \growthDirection{} in Ge, and marginally so in Si. In any case, further spin-dependent terms appear unless the \growthDirection{} is along a high-symmetry axis ([001] or [111] or their equivalents) \cite{trebin_quantum_1979, winkler_anisotropic_1996}. These additional terms will induce mixing, which we examine next.

\subsection{Approximately pure spinors}

\newcommand{\thetaPlus}{\theta_{\scriptscriptstyle \hspace{-0.05cm} +}}
\newcommand{\thetaTotal}{\theta_{\scriptscriptstyle \hspace{-0.0cm} c}}

One of the upshots of the technical analysis presented in App.~\ref{app:proofs} is that the terms appearing in Eq.~\eqref{eq:H0-delfectedGrowthDirection} are of two different types:
\begin{subequations}
\label{eq:termsTypes}
\begin{align}
&\left. \begin{tabular}{c} $\{J_\dotAxis{x},J_\dotAxis{z}\}$ \vspace{0.1cm} \\ $\{J_\dotAxis{y},J_\dotAxis{z}\}$ \vspace{0.1cm}
	\end{tabular} \right\} \mathrm{spinor\, rotation}, \,H_\mathrm{rot}\, , \label{eq:rotationTerms}\\
& \left. \begin{tabular}{c} $\{J_\dotAxis{x},J_\dotAxis{y}\}$ \vspace{0.1cm} \\ $J_\dotAxis{x}^2-J_\dotAxis{y}^2$ \vspace{0.1cm}
	\end{tabular} \right\}  \mathrm{spinor\, mixing}, \,H_\mathrm{mix}\, . \label{eq:mixingTerms}
\end{align}
\end{subequations}
When their amplitudes are small and in  leading order, the terms 
in Eq.~\eqref{eq:rotationTerms}
keep the hole spinors pure, in a conveniently redefined coordinate frame. On the other hand, the terms 
in Eq.~\eqref{eq:mixingTerms}
can not be accommodated as a reference frame rotation and they directly (in  leading order) decrease the spinor pureness, quantified using any of the measures $m_i$. The rotation effect of the first type of terms can be understood as a simple procedure of `completing the square',
\be
\label{eq:completingSquare}
J_\dotAxis{z}^2 + \epsilon \{J_\dotAxis{x},J_\dotAxis{z}\} = \left( \cos\thetaPlus J_\dotAxis{z} + \sin \thetaPlus J_\dotAxis{x}\right)^2 + O(\epsilon^2),
\ee
where $\sin \thetaPlus \sim \epsilon$. With this interpretation, the parameter $\thetaPlus$ is the angle by which the direction of the closest pure spinor deflects from the \growthDirection{}, and we call it `spinor deflection angle'. 

More precisely (see App.~\ref{app:proofs}),  the procedure in Eq.~\eqref{eq:completingSquare} corresponds to finding a coordinate frame where there are no rotation terms $H_\mathrm{rot}$. 
We find that in both [llh] and [0lh] scenarios the `$z$' axis of such suitably rotated coordinate frame, denoted by $\deflectedAxis{z}$, lies in the plane of rotation $\crystalAxis{z} \to \dotAxis{z}$ and can thus be parametrized simply.\footnote{This property does not hold in general, where one needs two parameters to specify the deflected axis direction. The simplification is another motivation for considering [llh] and [0lh] scenarios.}
With the \growthDirection{} parametrized by the last column of $R$ for angles $\phi, \theta, \phi^\prime$ as described by Eq.~\eqref{eq:coordinateRotation}, the axis $\deflectedAxis{z}$ is the last column of $R$ for angles 
$\phi, \theta+\thetaPlus, \phi^\prime$. 
 Solving Eq.~\eqref{eq:completingSquare} for Eq.~\eqref{eq:H0-firstScenario} and Eq.~\eqref{eq:H0-secondScenario}, we get analytical approximations,
\begin{subequations}
\label{eq:thetaPlus}
\begin{align}
\label{eq:thetaPlusllh}
&\mathrm{for\, [llh]: \quad} \sin \thetaPlus = \frac{\gamma_\delta}{8\overline{\gamma}}\left(2\sin2\theta +3\sin4\theta\right),\\
\label{eq:thetaPlus0lh}
&\mathrm{for\, [0lh]: \quad}\sin \thetaPlus = \frac{\gamma_\delta}{2\overline{\gamma}} \sin4\theta.
\end{align}
\end{subequations}
To arrive at these simple forms, we have neglected the angular dependences of the strength of the $J_\dotAxis{z}^2$ term by putting $\gamma_\delta=0$ for it.\footnote{Since we consider the zone-center Hamiltonian, all results derived from it should be understood as qualitative. Working with simplified analytical formulas is then justified. We give exact formulas in App.~\ref{app:exactFormulas}, see Eq.~\eqref{eq:thetaPlusExact}.} 
The formulas in Eq.~\eqref{eq:thetaPlus} give good approximation for the direction that can be interpreted as either: 1) the direction\footnote{Here by `direction of the spinor' we mean the direction defining the `$z$' axis in Eqs.~\eqref{eq:unperturbedSpinorH0}-\eqref{eq:pureHoleSpinors}.}  of pure spinors $\Psi_{\pm3/2}^\mathrm{pure}$ that are `closest' to the actual heavy-hole spinor subspace, or 2) the direction of the actual spinors $\Psi_{\pm 3/2}$. 
This interpretation is supported by the definitions of the measures $m_1$ and $m_2$.

\begin{figure}
\begin{center}
\hspace{-0.3cm}
\includegraphics[width=0.55\linewidth]{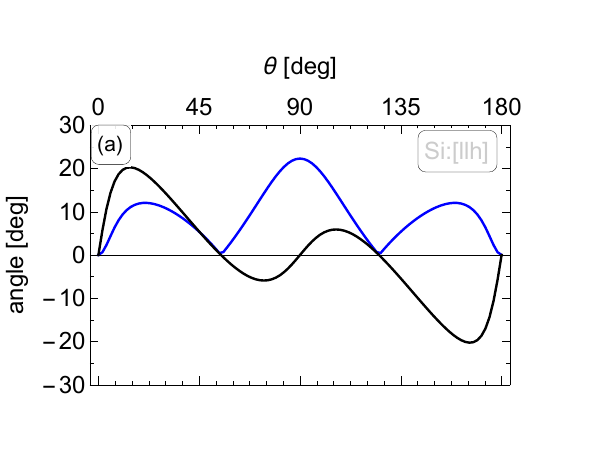}
\hspace{-1.2cm}
\includegraphics[width=0.55\linewidth]{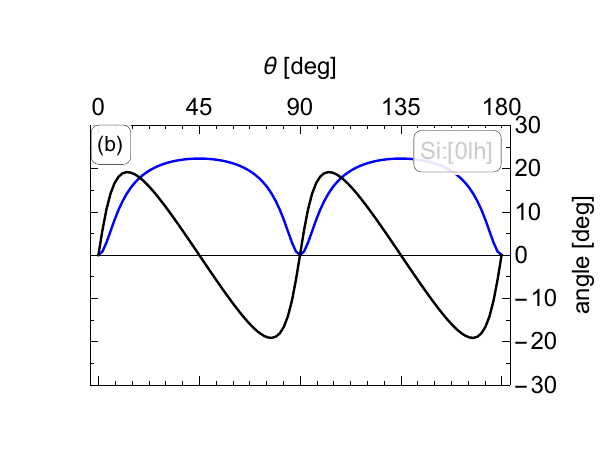}\\
\vspace{-0.5cm}
\includegraphics[width=0.2\linewidth]{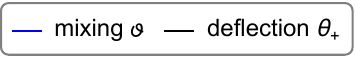}\\
\vspace{-0.5cm}
\hspace{-0.3cm}
\includegraphics[width=0.55\linewidth]{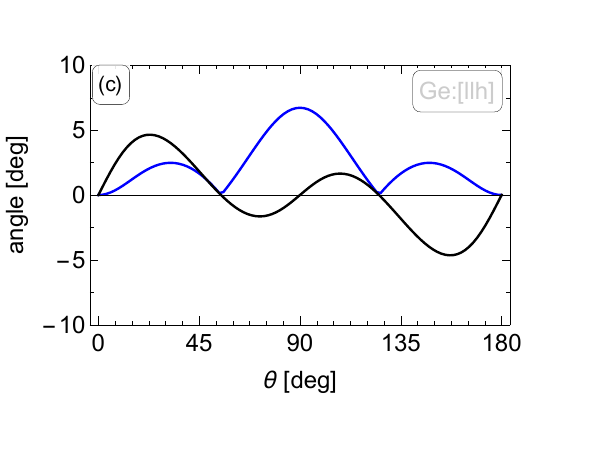}
\hspace{-1.2cm}
\includegraphics[width=0.55\linewidth]{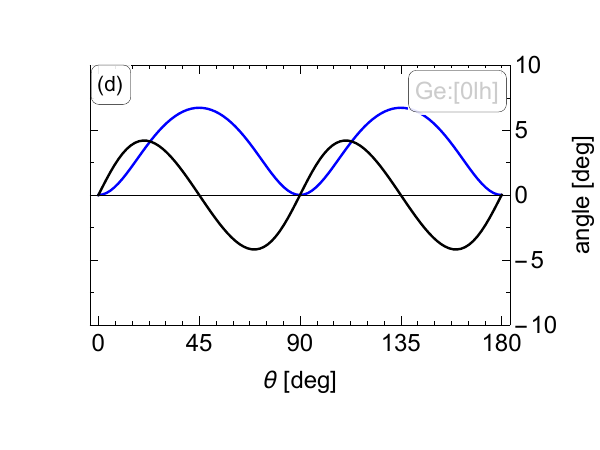}\\
\vspace{-1cm}
\end{center}
\caption{
\textbf{The mixing angle and the deflection angle} 
for (a) Si:[llh] (b) Si:[0lh] (c) Ge:[llh] (d) Si:[0lh]. 
The mixing angle $\vartheta$ (blue) is defined in Eq.~\eqref{eq:alpha} and the deflection angle (black) is calculated according to Eq.~\eqref{eq:thetaPlusExact}. The horizontal axis shows the angle $\theta$, which parametrizes the \growthDirection{} orientation as described in Sec.~\ref{sec:scenarios}.
\label{fig:pureHHmeasuresNEW}
}
\end{figure}

\subsection{Mixing degree}

The mixing and deflection angles are plotted in Fig.~\ref{fig:pureHHmeasuresNEW}. Starting with the [llh] scenario and Si parameters, Fig.~\ref{fig:pureHHmeasuresNEW}(a), we observe that as the \growthDirection{} rotates, the mixing angle varies between 0 and $20^\circ$, the maximum achieved at $\theta=\pi/2$ corresponding to [110]. 
The deflection angle varies by similar magnitudes, though it is not specified uniquely by the mixing strength: while both go to zero at high-symmetry growth directions such as [001] and [111], the deflection angle is zero also at [110] where the mixing is maximal. In this case, the Hamiltonian does not contain any deflection-generating terms $\{ J_{\dotAxis{x}/\dotAxis{y}},J_\dotAxis{z}\}$, but only the purely mixing term $\{ J_\dotAxis{x},J_\dotAxis{y}\}$. Looking at the [0lh] scenario shown in Fig.~\ref{fig:pureHHmeasuresNEW}(b), one can see that while the curves have different shapes, the overall magnitude of the effects is similar. Turning next to germanium, plotted in Fig.~\ref{fig:pureHHmeasuresNEW}(c)--(d), the curves have shapes qualitatively similar to those for silicon,
except for the overall magnitude. In Ge, both angles are much smaller. The difference can be traced back to a much stronger dominance of the term $-J_\dotAxis{z}^2$ in Ge than in Si (compare the black solid versus dashed lines in Fig.~\ref{fig:spinOperatorParts}).\footnote{We note in passing that a different energy distance to the split-off valence band leads to additional differences of silicon versus germanium concerning the spin structure of the valence band \cite{sarkar_electrical_2023,wang_electrical_2024}. The split-off band is not included in our model, which then does not contain these effects.}

\section{The form and strength of the generated SOI}

We now turn to $\mSOI$, motivated by an important practical question: Assuming that there is Rashba SOI in the bulk, Eq.~\eqref{eq:bulkRashba}, what is the emerging SOI in the heavy-hole subband? Apart from being expected in Si and Ge, the form of Eq.~\eqref{eq:bulkRashba} is beneficial due to its rotational symmetry.\footnote{Invariance with respect to rotating all three vectors $\mathbf{E}$, $\mathbf{J}$, and $\mathbf{k}$.} We will find that the SOI generated in the heavy-hole subband consists of two terms,
\be
\label{eq:SOI total}
H_r^\mathrm{\scriptscriptstyle 2D} =H_{r,\mathrm{def}}^\mathrm{\scriptscriptstyle 2D} +H_{r,\mathrm{mix}}^\mathrm{\scriptscriptstyle 2D},
\ee
induced, respectively, by the coordinate frame deflection and the heavy-hole--light-hole mixing. We next explain the origin of these two terms.

\subsection{SOI generated by deflection terms}

In the previous section, we have explained that terms in Eq.~\eqref{eq:rotationTerms} 
can be removed from the Hamiltonian by a suitable coordinate rotation. In this rotated frame, the leading-order term in the spinor Hamiltonian is $-J_\deflectedAxis{z}^2$. On the other hand, the electric field $\mathbf{E}$ inducing the bulk Rashba SOI  is along the \growthDirection{} $\dotAxis{z}$.\footnote{For us, this is actually the definition of the 2DHG normal: The 2DHG plane is the plane perpendicular to the value of the defining confinement field $\mathbf{E}$ at the location of the quantum dot.} The deflection of these two axes, by angle $\thetaPlus$, induces SOI in the heavy-hole subband. Indeed, using the pure-hole spinor projection rules Eq.~\eqref{eq:projectionRules0} in the deflected coordinate system one has (see App.~\ref{app:derivation})
\be
\label{eq:projectedRashba0}
H_{r,\mathrm{def}}^\mathrm{\scriptscriptstyle 2D}  =  \alpha_r^\mathrm{\scriptscriptstyle 3D}\,  \sin(\thetaPlus)\,	E_\dotAxis{z} \,s_\deflectedAxis{z} (\mathbf{k} \cdot \mathbf{m}),
\ee
where we recall that $\mathbf{m}$ is a unit vector that lies in the $\dotAxis{x}$$\dotAxis{y}$ plane and is perpendicular to the plane $\dotAxis{z}$$\deflectedAxis{z}$. We thus conclude that the deflection results in a SOI in the first order of $\thetaPlus$, in turn in the first order of the rotation-inducing strengths in Eq.~\eqref{eq:rotationTerms}. The SOI  is of unusual form, with an (almost) out-of-plane pseudospin operator $s_\deflectedAxis{z}$. Also, since it contains a single pseudospin component, the SOI field direction does not depend on the momentum direction. Such SOI generates a persistent (pseudo-) spin helix \cite{schliemann_nonballistic_2003, koralek_emergence_2009, dettwiler_stretchable_2017, schliemann_colloquium_2017, passmann_dynamical_2019}.

\subsection{SOI generated by mixing}

Since the Luttinger Hamiltonian is bilinear in spin operators, after removing the rotation-generating terms, there are only two remaining possibilities, given in Eq.~\eqref{eq:mixingTerms}.\footnote{\label{fnt:Jpm} In further derivations it is beneficial to use alternative combinations $J_\pm^2 = \left(J_\dotAxis{x}^2-J_\dotAxis{y}^2 \pm i \{J_\dotAxis{x}, J_\dotAxis{y}\}\right) /2$, where the raising and lowering operators are defined by $J_\pm = (J_x \pm i J_y)/\sqrt{2}$.} Assuming that such terms are present (or generated by the rotation into the canonical coordinate frame in second order in $\gamma_\delta$), we are interested in the degree of mixing that they result in, if quantified by the measure
proposed in Eq.~\eqref{eq:measure3}.

With this goal, we derive the effective SOI in the heavy-hole subspace that is induced by the bulk Rashba SOI, Eq.~\eqref{eq:bulkRashba}, and the heavy-hole--light-hole mixing terms in Eq.~\eqref{eq:mixingTerms} denoted as $H_\mathrm{mix}$. We assume that both of these terms are small so that we can treat them perturbatively. Using quasidegenerate perturbation theory,\footnote{See Footnote 1 in Ref.~\cite{stano_orbital_2019} for the nomenclature.} we get the effective Hamiltonian\footnote{The simple form of the effective interaction is yet another advantage of adopting the zone-center approximation. Aiming at an effective Hamiltonian in the first order in $H_r^\mathrm{\scriptscriptstyle 3D}$, we are effectively evaluating the matrix elements of the operators $\mathbf{J}$ in the subspace of the two heavy-hole spinors. These spinors are given solely by $H_S$, the orbital degrees of freedom in Eq.~\eqref{eq:unperturbedSpinorH0} are irrelevant.}
\be
\label{eq:effectiveHamiltonian}
H_r^\mathrm{\scriptscriptstyle 2D} =  P_\mathrm{hh} \left( H_r^\mathrm{\scriptscriptstyle 3D} -\frac{ \{H_r^\mathrm{\scriptscriptstyle 3D}, H_\mathrm{mix} \}}{\Delta}  \right) P_\mathrm{hh} + O(H_\mathrm{mix}^2).
\ee
Here, $P_\mathrm{hh}$ is the projector to the heavy-hole subspace [equal to $\rho_{\pm 3/2}$  defined below Eq.~\eqref{eq:measure2}] and the two terms in the bracket give the zeroth and first-order perturbation in $H_\mathrm{mix}$. 
The first term inside the large brackets in Eq.~\eqref{eq:effectiveHamiltonian} embodies the pure heavy-hole limit. Relying on the projection rules given in Eq.~\eqref{eq:projectionRules0}, without any deflection there is no SOI induced in the heavy-hole subspace. A finite deflection generates the expression given in Eq.~\eqref{eq:projectedRashba0}.  The second term in the large brackets in Eq.~\eqref{eq:effectiveHamiltonian} induces finite matrix elements of the in-plane spin operators, as $O(H_\mathrm{mix})$ corrections to Eq.~\eqref{eq:projectionRules0}. The explicit formulas are listed in Table \ref{tab:projectionRules}. 

\begin{table}
\begin{tabular}{@{\quad}c@{\quad}c@{\quad}c@{\qquad}c@{\qquad}c@{}}
\toprule
&&\multicolumn{3}{c}{$H_\mathrm{mix}$}\\
\cmidrule{3-5}
$H_r^\mathrm{\scriptscriptstyle 3D} $&&none&$\frac{c_{xy}}{2}\{J_x, J_y\}$ & $\frac{c_{xx}-c_{yy}}{2}(J_x^2 - J_y^2)$\\
\midrule
\addlinespace[0.5em]
$J_x$&\multirow{5}{*}{$\toHH$}&0& $-\frac{3c_{xy}}{\Delta} \times s_y $ & $ -\frac{3(c_{xx}-c_{yy})}{\Delta} \times s_x$\\
\addlinespace[-0.5em]\\
$J_y$ &&0& $\frac{3c_{xy}}{\Delta} \times s_x$ & $ -\frac{3(c_{xx}-c_{yy})}{\Delta} \times s_y$\\
\addlinespace[-0.25em]\\
$J_z$ &&$3s_z$&0&0\\
\addlinespace[0.3em]
\bottomrule
\end{tabular}
\caption{
\label{tab:projectionRules}
\textbf{Bulk-hole spin to heavy-hole pseudospin projection rules.} 
With heavy-hole--light-hole mixing-inducing terms $H_\mathrm{mix}$ present in the Hamiltonian, the bulk spin operator given in the first column induces finite matrix elements within the heavy-hole subspace, expressed using an effective pseudo-spin operator $\mathbf{s}$ given in the corresponding table entry. The coefficients $c_{ij}$ are the strengths of the terms in $H_\mathrm{mix}$, as defined in Eq.~\eqref{eq:H0-delfectedGrowthDirection}. The given projections are valid in any coordinate frame where the bulk hole Hamiltonian is $c_{zz}J_z^2+H_\mathrm{mix}$. In the text, this frame is denoted with doubly primed coordinates $\deflectedAxis{x}$, $\deflectedAxis{y}$,  $\deflectedAxis{z}$. However, we omit the primes in the table to ease the notation.}
\end{table}

\begin{figure}
\begin{center}
\hspace{-0.0cm}
\includegraphics[width=0.47\linewidth]{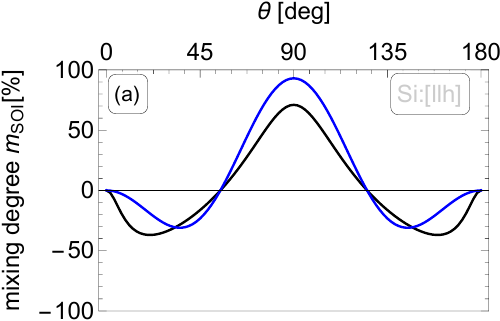}
\hspace{0.0cm}
\includegraphics[width=0.47\linewidth]{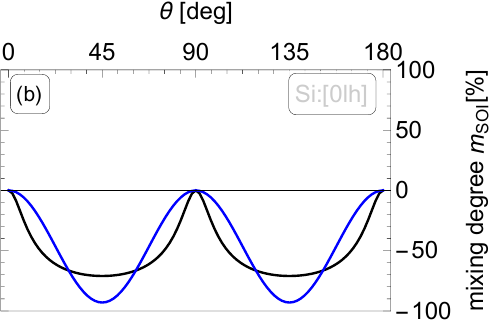}\\
\vspace{-0.0cm}
\includegraphics[width=0.35\linewidth]{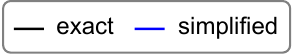}\\
\vspace{-0.0cm}
\hspace{-0.0cm}
\includegraphics[width=0.47\linewidth]{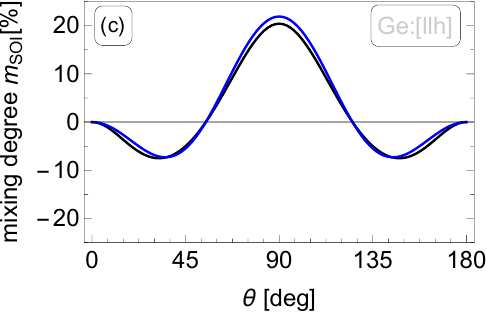}
\hspace{0.0cm}
\includegraphics[width=0.47\linewidth]{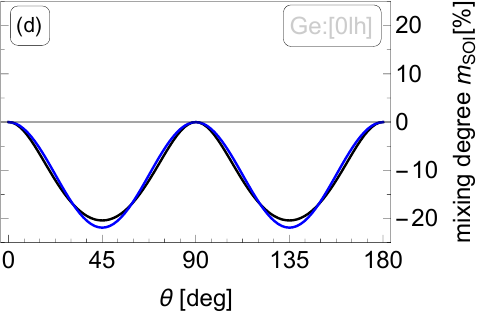}\\
\vspace{-0.3cm}
\end{center}
\caption{
\textbf{Generated SOI  strength}. The mixing strength $\mSOI$ for Si (upper row) and Ge (bottom row) for [llh] (left column) and  [0lh] (right column).  The black curves give Eq.~\eqref{eq:m3Exact}, the blue curves show the approximation in Eq.~\eqref{eq:measure3results}.
\label{fig:generatedSOC}
}
\end{figure}

The form and strength of the SOI  in the heavy-hole subband induced by the bulk Rashba SOI and light-hole--heavy-hole mixing can now be read off from Tab.~\ref{tab:projectionRules}. Assume that only one type of mixing term is present, and start with $H_\mathrm{mix} \propto J_\deflectedAxis{x}^2-J_\deflectedAxis{y}^2$. Using the last column of the table, one sees that the projection preserves the bulk SOI  form with a renormalized strength\footnote{We 
state the result mixing the singly and doubly primed coordinate frames: while the singly-primed ones are the natural frame for vectors $\mathbf{E}$ and $\mathbf{k}$, being the coordinate system of the confinement, the doubly-primed ones are natural for the spin. 
In Eq.~\eqref{eq:projectedRashba1}, one can replace $s_\deflectedAxis{x} \to s_\dotAxis{x}$ and $s_\deflectedAxis{y} \to s_\dotAxis{y}$. Within the precision of these formulas, set by the perturbation order $O(H_\mathrm{mix})$, the two sets of operators do not differ. See App.~\ref{app:derivation} for the derivation and comments.}
\be
\label{eq:projectedRashba1}
H_{r,\mathrm{mix}}^\mathrm{\scriptscriptstyle 2D}  = -3\frac{c_{\deflectedAxis{x}\deflectedAxis{x}}-c_{\deflectedAxis{y}\deflectedAxis{y}}}{\Delta} E_\dotAxis{z} (s_\deflectedAxis{x} k_\dotAxis{y}-s_\deflectedAxis{y} k_\dotAxis{x}).
\ee
For the other mixing term, $H_\mathrm{mix} \propto \{ J_\deflectedAxis{x}, J_\deflectedAxis{y} \}$, the  middle column of Table \ref{tab:projectionRules} gives
\be
\label{eq:projectedRashba2}
H_{r,\mathrm{mix}}^\mathrm{\scriptscriptstyle 2D}  = -3\frac{c_{\deflectedAxis{x}\deflectedAxis{y}}}{\Delta} E_\dotAxis{z} (s_\deflectedAxis{x} k_\dotAxis{x} + s_\deflectedAxis{y} k_\dotAxis{y}).
\ee
While this form looks nonstandard, it is unitarily equivalent to Eq.~\eqref{eq:projectedRashba1} upon rotating the pseudospin axes by $-\pi/2$ around $\deflectedAxis{z}$. When both mixing terms are present, their action in inducing terms in the effective Hamiltonian is additive, as follows from Eq.~\eqref{eq:effectiveHamiltonian}. The resulting interaction is a sum of the terms in Eq.~\eqref{eq:projectedRashba1} and Eq.~\eqref{eq:projectedRashba2}. Due to the $-\pi/2$ rotation in the second equation, the two SOI fields are orthogonal and adding them will result in an interaction that is unitarily equivalent to the standard Rashba SOI
\be
\label{eq:Hr2D}
H_{r,\mathrm{mix}}^\mathrm{\scriptscriptstyle 2D} = - \mSOI \times U_\varphi \alpha^\mathrm{\scriptscriptstyle 3D} E_\dotAxis{z} (s_\deflectedAxis{x} k_\dotAxis{y}-s_\deflectedAxis{y} k_\dotAxis{x}) U_\varphi^\dagger,
\ee
with the pseudospin rotation dependent on the strengths of the mixing terms, 
\begin{subequations}
\label{eq:varphi}
\begin{align}
U_\varphi&=\exp( -i s_\deflectedAxis{z} \varphi ),\\
\frac{\sin \varphi}{\cos\varphi} &= \frac{c_{\deflectedAxis{x}\deflectedAxis{y}}}{c_{\deflectedAxis{x}\deflectedAxis{x}}-c_{\deflectedAxis{y}\deflectedAxis{y}}},
\end{align}
\end{subequations}
and the strength-renormalization\footnote{With the definitions in Eq.~\eqref{eq:varphi} and Eq.~\eqref{eq:m3}, the sign of the measure $\mSOI$ is ambiguous, as a negative sign can be traded for a $\pi$ shift in the angle $\varphi$. When plotting Fig.~\ref{fig:generatedSOC}, we opt for a smooth curve, allowing $\mSOI$ to change sign, while keeping the function $\varphi(\theta)$ continuous, without any jumps by $\pi$.}
\be
\label{eq:m3}
|\mSOI| = 3\frac{\sqrt{(c_{\deflectedAxis{x}\deflectedAxis{x}}-c_{\deflectedAxis{y}\deflectedAxis{y}})^2+c_{\deflectedAxis{x}\deflectedAxis{y}}^2}}{\Delta}.
\ee
Inserting $\Delta = -2c_{\deflectedAxis{z}\deflectedAxis{z}}$ and using the definitions in App.~\ref{app:proofs}, we finally get
\be
\label{eq:m3Exact}
|\mSOI| = \sqrt{3} \tan \vartheta.
\ee
The last equation, the strength of the induced Rashba SOI in the heavy-hole subband appearing upon mixing, is the main result of this section. In a simple expression, it embodies the measure for the heavy-hole--light-hole mixing effects on SOI. 

We also provide simplified formulas for the SOI strength using the Hamiltonian expressed in the 2DHG coordinate frame. They follow from Eq.~\eqref{eq:m3} by putting $c_{\deflectedAxis{i}\deflectedAxis{j}} \approx c_{\dotAxis{i}\dotAxis{j}}$ and $\Delta \approx -2c_{\dotAxis{z}\dotAxis{z}} \approx 4 \overline{\gamma}$. For our two scenarios, we read off the coefficients $c_{\dotAxis{i}\dotAxis{j}}$ from Eq.~\eqref{eq:H0-delfectedGrowthDirection} and get\footnote{For the [0lh] scenario, the same angular dependence as in Eq.~\eqref{eq:measure3results-b} was derived in Ref.~\cite{budkin_spin_2022} for a spin-orbit interaction induced by a discontinuity of Luttinger parameters at the heterostructure interfaces.}
\begin{subequations}
\label{eq:measure3results}
\begin{align}
\label{eq:measure3results-a}
&\mathrm{for\,[llh]:}\quad |\mSOI|=\frac{ 2 \gamma_\delta \sin^2\theta (3\cos2\theta + 1)}{\Delta},\\
\label{eq:measure3results-b}
&\mathrm{for\,[0lh]:}\quad |\mSOI|=\frac{ \gamma_\delta \sin^2 (2\theta)}{\Delta}.
\end{align}
\end{subequations}
We plot $\mSOI$ in Fig.~\ref{fig:generatedSOC}. The exact results are in black, the simplified ones in blue. Again, the leading-order approximations are excellent for Ge, while somewhat larger discrepancies are visible for Si. In Si, the strength of the induced SOI can reach above 50\% of the bulk interaction strength, while in Ge, the ratio can be up to 20\%.

\section{Applications}

Our results allow for a qualitative understanding of the spin structure of holes in quasi-two-dimensional confinement and structures derived from it, such as split-gate QPC or planar quantum dots.\footnote{We remind that our approach only grasps the effects included in the zone-center Hamiltonian. Especially, influences from the quantum-dot in-plane confinement (quantum-dot shape, squeezing, driving, and similar), will induce additional effects through the in-plane momenta part of the Luttinger Hamiltonian that we have neglected by adopting Eq.~\eqref{eq:unperturbedSpinorH0}.} We demonstrate the applications of the results with two examples. First, we look at the effective Hamiltonian that describes a heavy hole spin in a finite magnetic field and analyze the associated $g$ tensor. Second, we consider the effects of strain and its gradients.

\subsection{Zeeman interaction and $g$ tensor}

\label{sec:4A}

We now consider the case where a magnetic field $\mathbf{B}$ is applied.\footnote{While \label{fnt:ZeemanTRS} this term breaks the TRS, we include it perturbatively. The heavy-hole subspace is defined from a TRS-preserving Hamiltonian, as pointed out in Footnote \ref{fnt:TRS}.}
We ignore the orbital effects and neglect the small cubic Zeeman term\footnote{We give the matrix elements of the cubic Zeeman term in App.~\ref{app:cubic} for completeness.}.
The corresponding bulk Hamiltonian is 
\begin{equation}
\label{eq:Hz3D}
H_\mathrm{z}^\mathrm{\scriptscriptstyle 3D}  = 2\mu_B \kappa \mathbf{B} \cdot \mathbf{J},
\end{equation}
where $\mu_B$ is the Bohr magneton, $\kappa$ is the $g$ factor. The bulk interaction induces an effective Hamiltonian within the heavy-hole subspace. We can get it from Eq.~\eqref{eq:effectiveHamiltonian} by the substitution $H_\mathrm{r}^\mathrm{\scriptscriptstyle 3D} \to H_\mathrm{z}^\mathrm{\scriptscriptstyle 3D}$. Together with Tab.~\ref{tab:projectionRules}, one immediately gets\footnote{A notation-related comment: we do not use any explicit sign for the multiplication of vectors and matrices. The vectors, such as $\mathbf{B}$ or $\mathbf{J}$ are column vectors. They become row vectors upon transpose, for example, $\mathbf{B}^T$ or $\mathbf{J}^T$. The only explicit sign concerning tensor products that we use is the scalar product sign $\cdot$, which removes the need for a transpose, for example, $\mathbf{B} \cdot \mathbf{J} \equiv \mathbf{B}^T \mathbf{J}$. It has lower precedence than a tensor multiplication without a sign; the right-hand sign of Eq.~\eqref{eq:Hz2D} with the operator precedence made explicit by brackets is $(\mu_B \mathbf{B})\cdot (\mathbf{g} U_\varphi \mathbf{s} U^\dagger_\varphi)$.}
\begin{equation}
\label{eq:Hz2D}
H_\mathrm{z}^\mathrm{\scriptscriptstyle 2D} = \mu_B \mathbf{B} \cdot \mathbf{g} U_\varphi \mathbf{s} U^\dagger_\varphi,
\end{equation}
where $U_\varphi$ is defined in Eq.~\eqref{eq:varphi} and $\mathbf{g}$ is the $g$ tensor. Applying Tab.~\ref{tab:projectionRules} means we have derived the preceding equation in the doubly primed coordinate system. Here, the $g$ tensor is diagonal with the following components  
\begin{equation}
\label{eq:gTensorResult}
\mathbf{g}|_{\deflectedAxis{x}\deflectedAxis{y}\deflectedAxis{z}}=
 \mathrm{diag}(2\kappa\,\mSOI, 2\kappa\,\mSOI, 6\kappa).
\end{equation}
In this convenient coordinate system (the canonical frame), the effects of the heavy-hole--light-hole mixing on the $g$ tensor are simple. At zero mixing, the $g$-tensor components in the $\deflectedAxis{x}\deflectedAxis{y}$ plane are zero. Nonzero mixing induces nonzero and isotropic $g$ tensor in this plane. This simplicity is obscured by the fact that the normal of this plane is rotated with respect to the \growthDirection{}. The $g$-tensor matrix for the magnetic field components evaluated in the other two coordinate systems is 
\begin{subequations}
\label{eq:gTensorTransformations}
\begin{align}
\mathbf{g}|_{\dotAxis{x}\dotAxis{y}\dotAxis{z}}&=R[\phi, \thetaPlus,-\phi] \,\, \mathbf{g}|_{\deflectedAxis{x}\deflectedAxis{y}\deflectedAxis{z}}
,\\
\mathbf{g}|_{\crystalAxis{x}\crystalAxis{y}\crystalAxis{z}}&=R[\phi, \theta,-\phi] \,\, \mathbf{g}|_{\dotAxis{x}\dotAxis{y}\dotAxis{z}} %
,
\end{align}
\end{subequations}
with $\phi=\pi/4$ for [llh] and $\phi=\pi/2$ for [0lh].\footnote{We remark that $g$ tensors in 2DHG with arbitrary growth directions were investigated in Ref.~\cite{qvist_anisotropic_2021}.}

\begin{figure}
\begin{center}
\includegraphics[width=0.9\linewidth]{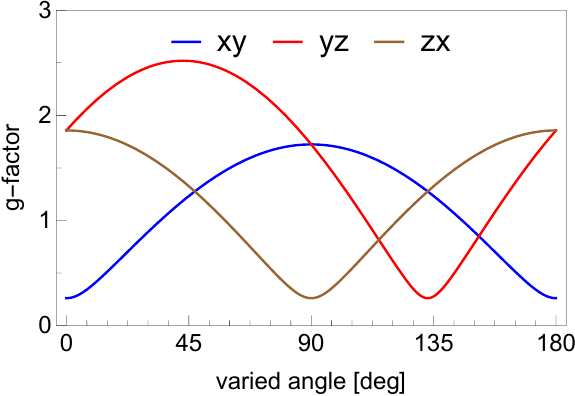}\\
\end{center}
\caption{
$\boldsymbol{g}$ \textbf{factor} as a function of the magnetic field direction $\mathbf{B}=B(\cos\alpha \sin\beta, \sin\alpha\sin\beta, \cos\beta)$ varied over a semicircle in three orthogonal planes as denoted by the legend: $\dotAxis{x}\dotAxis{y}$ means $\beta=\pi/2$ and $\alpha$ is varied, $\dotAxis{y}\dotAxis{z}$ means $\alpha=0$ and $\beta$ is varied, and $\dotAxis{z}\dotAxis{x}$ means $\alpha=\pi/2$ and $\beta$ is varied. The varied angle is on the horizontal axis of the figure. 
We took the parameters for Si, scenario $[0lh]$ with $\theta=30^\circ$, and evaluated the $g$ factor as $g=|\mathbf{B}^T \mathbf{g}|/B$ using Eqs.~\eqref{eq:thetaPlus}, \eqref{eq:gTensorResult}, and \eqref{eq:gTensorTransformations}.
\label{fig:gFactor}
}
\end{figure}

As an illustration, we plot the $g$ factor, defined as the norm of the vector $\mathbf{B}^T \mathbf{g}/B$, calculated using Eq.~\eqref{eq:gTensorResult}, \eqref{eq:gTensorTransformations}, and the simplified expression for the $\mSOI$, Eq.~\eqref{eq:thetaPlus}, in Fig.~\ref{fig:gFactor}. We find it remarkable that our simple model shows $g$-factor dependences qualitatively similar to those observed experimentally in an MOS Si quantum dot as well as in much more elaborate simulations \cite{liles_electrical_2021}.\footnote{For example, we notice the nonsinusoidal shape (sharper minima versus broader maxima) of some of the curves in Fig.~\ref{fig:gFactor} and in Ref.~\cite{liles_electrical_2021}. Qualitatively similar curves have been obtained in a Si finFET quantum dot, in measurements accompanying Ref.~\cite{camenzind_spin_2022} (L. Camenzind and A. Kuhlmann, private communication; and Fig.~5.1 and Fig.~A.21 in Ref.~\cite{geyer_spin_2023}).}

\subsection{Spin qubit: dephasing and Rabi driving}

The $g$ tensor derived in the previous section can be used to analyze a heavy-hole spin qubit. We start by rewriting Eq.~\eqref{eq:Hz2D} as
\begin{equation}
\label{eq:effectiveH}
H_q = \mu_B [R_\varphi (\mathbf{B}^T \mathbf{g} )] \cdot \mathbf{s} \equiv \mu_B \mathbf{B}_q \cdot \mathbf{s},
\end{equation}
where $R_\varphi$ induces the same rotation (around $\deflectedAxis{z}$ by angle $\varphi$) on  three-dimensional vectors as $U_\varphi$ induces on spinors [see Eq.~\eqref{eq:RU_correspondence}]. We have also introduced the effective magnetic field $\mathbf{B}_q$ coupled to the pseudospin $\mathbf{s}$.

Several standard performance metrics, including the qubit dephasing, lifetime, or Rabi frequency, can be extracted from the effective spin-qubit Hamiltonian in Eq.~\eqref{eq:effectiveH}. They follow from its dependence on electric and magnetic fields, either controlled or fluctuating due to noise. We will not go into quantitative details since our model omits details of the quantum-dot confinement. However, we can use Eq.~\eqref{eq:effectiveH} to elucidate the physical origins of some of these effects. To this end, we emulate confinement changes as changes of the parameter $\theta$. We recall the discussion in the first paragraph of Sec.~\ref{sec:deflection}, that the \growthDirection{} is influenced by gates, for example, by pulling the confined hole against an interface or an impurity. 

With this interpretation of $\theta$, we start by the examination of `sweet spots', which are directions of the magnetic field $\mathbf{B}$ where the $g$ factor achieves its extremum with respect to variations in $\theta$. Since $\theta$ is an angle, the $g$ factor curve must be periodic in it and there will be at least two sweet spots (in $\theta$) for any fixed $\mathbf{B}$. Translating into equations, one would look for solutions of the following requirement:
\begin{subequations}
\label{eq:metrics}
\begin{equation}
\label{eq:T2}
\mathrm{maximal\, T_2^*}: \quad \min_{\mathbf{B}} |\mathbf{B}_q \cdot \partial_\theta \mathbf{B}_q|.
\end{equation}
Similarly, the qubit relaxation is mediated through the transverse matrix element of the effective magnetic field,
\begin{equation}
\label{eq:T1}
\mathrm{maximal\, T_1}: \quad \min_{\mathbf{B}} |\mathbf{B}_q \times \partial_\theta \mathbf{B}_q|.
\end{equation}
Since usually the relaxation is slow and not of concern, instead of minimizing this matrix element to minimize the relaxation rate, one is interested in finding its maximum, as the same matrix element mediates qubit Rabi rotations (oscillations) under resonant excitation. To describe such a situation, one needs to include the effects which originate in the time dependence of the frame in which the effective Hamiltonians, Eqs.~\eqref{eq:effectiveHamiltonian}, \eqref{eq:Hr2D}, \eqref{eq:Hz2D} were derived. As we show in App.~\ref{app:Heff}, the time dependence generates an additional effective time-dependent magnetic field 
\begin{equation}
\label{eq:BU}
\mathbf{B}_U  =  -\mathbf{m} \frac{\hbar}{2 \kappa \mu_B} \partial_t (\theta+\thetaPlus).
\end{equation}
\end{subequations}
Upon adding $\mathbf{B}_U$ to $\mathbf{B}_q$, one can, for example, search for the maximum of the right-hand side of Eq.~\eqref{eq:T1}.
As stated, we could straightforwardly plot any of these matrix elements in the same way as the $g$ factor in Fig.~\ref{fig:gFactor} and examine the extrema. However, since there have been several recent works that do such an analysis \cite{bosco_hole_2021,wang_optimal_2021,malkoc_charge-noise-induced_2022,michal_tunable_2023}, instead of repeating it, we comment on the sources of Rabi oscillations appearing in our model. 

We consider that an oscillating gate potential induces small oscillations in $\theta$. The corresponding time-dependent effective Hamiltonian is 
\begin{equation}
\label{eq:Rabi}
H_q  =  \mu_B \left( \mathbf{B} + \mathbf{B}_U \right) \cdot \mathbf{g} U_\varphi \mathbf{s} U^\dagger_\varphi.
\end{equation}
Varying $\theta$ will induce 1) changes of the $g$-tensor `in-plane' components through changing the value of $\mSOI$, 2)  changes of the value of the angle $\varphi$ defining the `in-plane' axes in the pseudo-spin space, 3) a fictitious magnetic field $\mathbf{B}_U$ with an oscillatory magnitude and fixed direction (along $\mathbf{m}$) due to the time-dependent frame rotation, and 4) oscillations of $\mathbf{B}$ (small harmonic displacements $\delta \mathbf{B}(t) \propto \mathbf{B} \times \mathbf{m}$), also due to the time dependence of the new reference frame. To draw analogies to the existing results for holes and electrons, we note that with the magnetic field applied in the 2DHG plane, the arising Hamiltonian terms are (approximately, neglecting here the deflection angle $\thetaPlus$) purely longitudinal in 1) and purely transverse in 2,3,4).\footnote{For channels 3 and 4, this is so if the in-plane field $\mathbf{B}$ is applied perpendicular to $\mathbf{m}$, in which case the (fictitious) magnetic fields $\mathbf{B}_U$ and $\delta \mathbf{B}(t)$ are both perpendicular to $\mathbf{B}$.} The latter channels will, therefore, be more efficient in inducing Rabi oscillations. Loosely speaking, channel 1) corresponds to the modulation of the $g$-tensor eigenvalues and 2-4) its eigenvectors (`iso-Zeeman' in the nomenclature of Ref.~\cite{crippa_electrical_2018}).
We point out that the Rabi oscillations associated with $H_q$ do not rely on the SOI terms in Eq.~\eqref{eq:SOI total}. Adding them to the effective Hamiltonian and promoting the in-plane momenta to $c$-numbers oscillating in time, 5) an additional channel to induce Rabi oscillation arises \cite{golovach_electric-dipole-induced_2006}.
 Since the linear-in-momenta SOI terms can be viewed, in the lowest order of the dot size over the SOI length, as a gauge transformation
\cite{braun_berrys_1996, aleiner_spin-orbit_2001, levitov_dynamical_2003,braunecker_spin-selective_2010}, this channel is similar in spirit to the fictitious magnetic field $\mathbf{B}_U$ induced by the time-dependent frame. In a given device all five channels will interfere in inducing Rabi oscillations.

\subsection{Strain and SOI}

We now look at strain, as it profoundly affects holes \cite{bir_symmetry_1974, winkler_spin-orbit_2003, thompson_uniaxial-process-induced_2006, sun_physics_2007}. The strain in the device is parametrized by a symmetric strain tensor $\epsilon_{ij}$. Its elements generate additional terms in the bulk hole Hamiltonian \cite{bir_symmetry_1974},
\begin{equation}
\label{eq:BirPikus1}
	\begin{split}
		H_{BP}  = & - \frac{2}{3} D_u \left( \epsilon_{xx} J_x^2+\mathrm{c.p.}\right) \\
		&-\frac{2}{3} D_u^\prime( \epsilon_{xy}\{J_x,J_y\} +\mathrm{c.p.}),
	\end{split}
\end{equation}
where the deformation potentials $D_u$ and $D_u^\prime$ are material parameters. 
Different from previous sections, we do not consider rotations of the \growthDirection{}. Instead, we consider various forms of the strain tensor, perhaps due to nearby gates, with consequences analogous to rotations. We assume a fixed \growthDirection{} [001] and consider in-plane strain $\epsilon_{xx}=\epsilon_{yy}=\epsilon_{||}<0$, relevant, for example, for lattice-matched heterostructures. The in-plane compression induces out-of-plane expansion $\epsilon_{zz}=-2\epsilon_{||}c_{12}/c_{11}$, with material-dependent elastic constants $c_{11}$ and $c_{12}$. Inserting this form of the diagonal strain-tensor components into Eq.~\eqref{eq:BirPikus1}, we get (constant omitted)
\begin{equation}
\label{eq:BirPikus2}
	\begin{split}
		H_{BP} = & - \frac{2}{3} D_u J_z^2 \left( \epsilon_{zz}-\epsilon_{||} \right) \\
		&
		-\frac{2}{3} D_u^\prime( \epsilon_{xy}\{J_x,J_y\} +\epsilon_{yz}\{J_y,J_z\}+\epsilon_{zx}\{J_z,J_x\}). 
	\end{split}
\end{equation}
The Hamiltonian is analogous to Eq.~\eqref{eq:H0-delfectedGrowthDirection}. 
The term in the first line sets the unperturbed system with the pure spinors given in Eq.~\eqref{eq:pureHoleSpinors} as its eigenstates and providing for the heavy-hole--light-hole splitting\footnote{
It also means that with $\epsilon_{xx}=\epsilon_{yy}$ and without any off-diagonal strain components, the strain does not induce heavy-hole--light-hole mixing \cite{lee_effects_1988}.
},
\begin{equation}
\label{eq:strain-Delta}
\Delta = \frac{4}{3} D_u \left( \epsilon_{zz} - \epsilon_{||} \right).
\end{equation}
The second line of Eq.~\eqref{eq:BirPikus2}
contains the perturbing terms. Classifying them using Eq.~\eqref{eq:termsTypes}, the last two terms induce deflection of the spinor axis away from the \growthDirection{}, here equal to the crystal axis $\dotAxis{z}=\crystalAxis{z}$. 
In analogy to Eq.~\eqref{eq:thetaPlus}, these off-diagonal strain components, quantified by $\epsilon_\mathrm{rot}^2 = \epsilon_{zx}^2+\epsilon_{yz}^2$, generate a deflection angle
\begin{equation}
\label{eq:strain-deflection}
\sin \thetaPlus = \frac{D_u^\prime}{D_u} \frac{\epsilon_\mathrm{rot} }{\epsilon_{zz} - \epsilon_{||}}.
\end{equation}
Neglecting the small contribution from this deflection, the heavy-hole--light-hole mixing is given by the first term in the second line of Eq.~\eqref{eq:BirPikus2},\footnote{In writing Eqs.~\eqref{eq:strain-Delta}, \eqref{eq:strain-cxy}, and \eqref{eq:strain-m3}, we are considering the strain effects in isolation from the orbital effects, neglecting the latter here. In reality, both orbital and strain effects contribute. For example, the heavy-hole--light-hole splitting is a sum of Eq.~\eqref{eq:Delta} (supplemented by the proper orbital-energy scale) and Eq.~\eqref{eq:strain-Delta}.}
\begin{equation}
\label{eq:strain-cxy}
c_{\deflectedAxis{x}\deflectedAxis{y}} \approx c_{\dotAxis{x}\dotAxis{y}} = c_{\crystalAxis{x}\crystalAxis{y}} = -\frac{4}{3} D_u^\prime \epsilon_{xy}.
\end{equation}
Using Eq.~\eqref{eq:strain-Delta}, we can evaluate Eq.~\eqref{eq:measure3},
\begin{equation}
\label{eq:strain-m3}
\mSOI = 3 \frac{D_u^\prime}{D_u} \frac{\epsilon_{xy}}{\epsilon_{zz} - \epsilon_{||}}.
\end{equation}

\begin{figure}
\begin{center}
\includegraphics[width=0.99\linewidth]{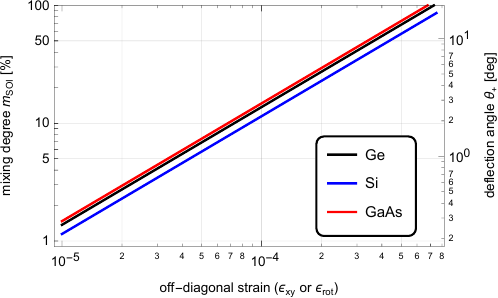}\\
\end{center}
\caption{
\textbf{Strain-generated mixing and deflection angle}. Referring to the left $y$ axis and the parameter $\epsilon_{xy}$ on the horizontal axis, the curves show Eq.~\eqref{eq:strain-m3}, the mixing measure $\mSOI$. Referring to the 
right $y$ axis and the parameter $\epsilon_\mathrm{rot}$ on the horizontal axis, the same curves show Eq.~\eqref{eq:strain-deflection}, the deflection angle $\thetaPlus$. We use $\epsilon_{||}=-0.61\%$, inspired by values appropriate for $\mathrm{Ge/Ge_{0.8}Si_{0.2}}$ heterostructures \cite{sammak_shallow_2019}. 
\label{fig:strainGenerated}
}
\end{figure}

The two quantities characterizing the perturbing terms, the mixing $\mSOI$ and the deflection angle $\thetaPlus$, are plotted in Fig.~\ref{fig:strainGenerated}. Once again, one can see a substantial SOI (measured by $\mSOI$; the left $y$ axis) arises already at relatively small strains. It is generated primarily by the in-plane off-diagonal strain components $\epsilon_{xy}$. With strains as small as a few times $10^{-4}$ (expected in typical devices \cite{corley-wiciak_nanoscale_2023}), the SOI that is generated in the heavy-hole subband has strength comparable to its value in the bulk ($\mSOI$ of order 1). The off-diagonal out-of-plane strain tensor elements, on the other hand, induce primarily a deflection of the pure spinor direction. The values denoted on the right $y$ axis convert $\mSOI$ to the corresponding deflection angle $\thetaPlus$. In the range plotted, the right-hand side of Eq.~\eqref{eq:strain-deflection} is small, the deflection angle $\thetaPlus$ is linear in the off-diagonal strain components, and the two quantities are proportional.%

\subsection{Strain gradients and Rabi driving}

Elaborate numerical and analytical analysis of Ref.~\cite{abadillo-uriel_hole-spin_2023} considered strain gradients, which together with periodically displacing a Ge quantum-dot hole in space induce Rabi oscillations. It was found that strain components $\epsilon_{xz}$ and $\epsilon_{yz}$ are most effective in inducing Rabi oscillations. We can provide a simple explanation based on our analysis. According to Eq.~\eqref{eq:BirPikus2}, the associated spin operators belong to $H_\mathrm{rot}$ in Eq~\eqref{eq:termsTypes}. Such operators lead, predominantly, to a rotation of the heavy-hole spinor (a `deflection' in our nomenclature). The corresponding rotation of the reference frame (see App.~\ref{app:Heff}) induces Rabi oscillations through the last term in Eq.~\eqref{eq:Rabi}, which is linear in $\partial_t \thetaPlus$, in turn linear in $\epsilon_\mathrm{rot}$ according to Eq.~\eqref{eq:strain-deflection}. The requirement for resonance means the frequency of driving equals the Zeeman energy $|\mu_B \mathbf{B}_q|.$ On the other hand, the term $\epsilon_{xy} \{J_x ,J_y\}$ belongs to $H_\mathrm{mix}$, meaning $\thetaPlus \approx 0$, and enters into Eq.~\eqref{eq:Rabi} in the second order, the second term in the bracket in Eq.~\eqref{eq:effectiveHamiltonian}. Its contribution to Rabi driving is thus expected to be smaller.

\section{Conclusions}

We have analyzed the heavy-hole--light-hole mixing in 2DHG. This mixing is central to many spin effects in devices based on holes in 2DHG. Particularly, it is responsible for the response of the hole spin to electric fields.
We have identified a canonical coordinate reference frame and a canonical mixing measure, the heavy-hole--light-hole mixing angle $\vartheta$. We have examined several derived measures of the mixing motivated by experiments and theory and quantified their relations to the mixing angle. We find that the measures based on the pure light-hole content in the wave function somewhat underestimate the efficiency of mixing in inducing the SOI: with a pure light-hole content that might be considered `small',\footnote{For example, Ref.~\cite{luo_discovery_2010} considers a 90\% content of heavy-hole as ``small" heavy-hole--light-hole mixing. Similarly, Ref.~\cite{winkler_anisotropic_1996} considers a 10\% light-hole admixture as ``small". 
}
the arising SOI is essentially of the same strength as the one in the bulk.

We have adopted the quasi-two-dimensional and zone-center approximation, which amounts to neglecting the in-plane momenta $k_\dotAxis{x} = 0 = k_\dotAxis{y}$. The approximation allows one to derive analytical results and get physical insight into various aspects of heavy-hole spins. Within this calculation scheme, we could identify different types of terms in the hole kinetic energy and strain Hamiltonians. Terms of the first type induce a rotation of the direction of spinors while keeping them pure. These terms are most efficient in inducing Rabi oscillations, acting analogously to Rashba or Dresselhaus SOIs inducing Rabi rotations in the conduction band through EDSR.
They also preserve the `$z$'  eigenvalues. Terms of the second type admix the pure light-hole components. They are less efficient for Rabi driving and more profoundly change the $g$ tensor. They are responsible for the $g$ tensor finite in-plane components.

The zone-center description provides a simple analytical model with which the basic characteristics of a spin qubit can be investigated without any sophisticated numerics or simulations. We have exemplified it by plotting the hole $g$ factor and gave formulas from which matrix elements responsible for hole spin dephasing, relaxation, or Rabi frequency can be obtained straightforwardly. An interesting extension of our work would be to include the in-plane orbital degrees of freedom perturbatively and analyze to which extent the simple scheme presented here can be used semiquantitatively. We leave the establishment of such a connection to elaborate numerical models for the future.

As a final comment, though we focus here on 2DHG, the assignment of the canonical coordinate system and the mixing angle apply to any spin-3/2 Hamiltonian with TRS. Thus, the mixing quantification can be applied, for example, to the effective Hamiltonian of a spin qubit that was obtained by integrating out the orbital degrees of freedom. While such an effective model is usually derived only for the ground state, that is, $2\times 2$ Hamiltonian, the calculations could be extended to cover the lowest $4 \times 4$ space. With this proviso, our results are applicable to a broader class of systems, to characterize the mixing of two Kramers-partner pairs.

\acknowledgments

We thank L. Camenzind, A. Kuhlmann, and D. Zumb\"uhl for useful discussions. We acknowledge the financial support from CREST JST Grant No.~JPMJCR1675 and from the Swiss National Science Foundation and NCCR SPIN Grant No.~51NF40-180604. P.S.~acknowledges the funding within the QuantERA II Programme that has received funding from the EU’s H2020 research and innovation programme under the GA No 101017733, and funding by Vedeck\'{a} Grantov\'{a} Agent\'{u}ra M\v{S}VVa\v{S} SR and SAV (VEGA Grant No.~2/0156/22).

\appendix

\section{TRS Hamiltonian for spin 3/2 in a simple form}

\label{app:proofs}

Here, we examine the meaning of the matrix $C$ of Hamiltonian coefficients $c_{ij}$ defined in Eq.~\eqref{eq:H0-delfectedGrowthDirection-coefficientsC} and its relation to the task of finding the eigenstates of the spin-3/2 Hamiltonian. We also elucidate the meaning of orthogonal transformations of the matrix $C$.

We consider a spin-3/2 Hamiltonian with the time-reversal symmetry (TRS). It is a four-by-four matrix that can be built from the following six linearly independent operators,
\be
\label{eq:TRS_basis}
\mathbb{1}, \, J_\dotAxis{z}^2, \, J_\dotAxis{x}^2-J_\dotAxis{y}^2, \, \{J_\dotAxis{x}, J_\dotAxis{y}\}, \,\{J_\dotAxis{y}, J_\dotAxis{z}\}, \,\{J_\dotAxis{z}, J_\dotAxis{x}\}.
\ee 
The basis choice implies the parameterization
\be
\label{eq:HC correspondence}
H = c_\dotAxis{\mathbb{1}} \mathbb{1} + \mathbf{J}^T \dotAxis{C} \mathbf{J},
\ee
that is, the constant $c_\dotAxis{\mathbb{1}}$ and a real symmetric matrix $\dotAxis{C}$
\be
\label{eq:gaugeDOFcomment}
\begin{tabular}{c|c}
\begin{tabular}{c}
$\scriptstyle{J_\dotAxis{x}}$ \\$\scriptstyle{J_\dotAxis{y}}$\\$\scriptstyle{J_\dotAxis{z}}$
\end{tabular} 
&
$\left(
\begin{tabular}{ccc}
$c_{\dotAxis{x}\dotAxis{x}}$ & $c_{\dotAxis{x}\dotAxis{y}}$ & $c_{\dotAxis{x}\dotAxis{z}}$\\
$c_{\dotAxis{x}\dotAxis{y}}$ & $c_{\dotAxis{y}\dotAxis{y}}$ & $c_{\dotAxis{y}\dotAxis{z}}$\\
$c_{\dotAxis{x}\dotAxis{z}}$ & $c_{\dotAxis{y}\dotAxis{z}}$ & $c_{\dotAxis{z}\dotAxis{z}}$
\end{tabular} 
\right).$
\end{tabular}
\ee
We remark here that Eq.~\eqref{eq:HC correspondence} contains a gauge degree of freedom, due to the relation $\mathbf{J}\cdot\mathbf{J}=\mathrm{const}$. If the matrix $\dotAxis{C}$ encodes only the non-constant terms from the basis in Eq.~\eqref{eq:TRS_basis}, the diagonal elements have to fulfill $c_{\dotAxis{x}\dotAxis{x}} + c_{\dotAxis{y}\dotAxis{y}} =0$. A generic matrix $\dotAxis{C}$ can be brought to this form by subtracting $(c_{\dotAxis{x}\dotAxis{x}} + c_{\dotAxis{y}\dotAxis{y}})/2$ from the diagonal and assigning this constant to $c_\dotAxis{\mathbb{1}}$. We adopt this definition of $c_\dotAxis{\mathbb{1}}$ (the gauge choice).

In the main text, we claim that it is beneficial to choose a coordinate system in which the Hamiltonian does not contain the last two terms of Eq.~\eqref{eq:TRS_basis}. We first show that such a Hamiltonian can be easily diagonalized. Omitting the constant and dividing the Hamiltonian by the factor $-c_{\deflectedAxis{z}\deflectedAxis{z}}$, we obtain ($\mathcal{H}=-H_S/c_{\deflectedAxis{z}\deflectedAxis{z}}$)
\be
\mathcal{H} = -J_\deflectedAxis{z}^2 +A (J_\deflectedAxis{x}^2-J_\deflectedAxis{y}^2) + B \{J_\deflectedAxis{x}, J_\deflectedAxis{y}\},
\ee
where the real constants are $A=(c_{\deflectedAxis{y}\deflectedAxis{y}}-c_{\deflectedAxis{x}\deflectedAxis{x}})/2c_{\deflectedAxis{z}\deflectedAxis{z}}$, $B=-c_{\deflectedAxis{x}\deflectedAxis{y}}/2c_{\deflectedAxis{z}\deflectedAxis{z}}$. Using the ladder operators (see Footnote \ref{fnt:Jpm}), we write this as
\be
\label{eq:H_s}
\mathcal{H} = -J_\deflectedAxis{z}^2 +\frac{\alpha}{\sqrt{3}}  J_\deflectedAxis{+}^2 +\frac{\alpha^*}{\sqrt{3}}  J_\deflectedAxis{-}^2,
\ee
where we introduced another (complex) constant $\alpha = \sqrt{3} (A- i B)$.
In the basis of pure heavy- and light-hole states, Eq.~\eqref{eq:pureHoleSpinors}, the Hamiltonian is block diagonal and has a single nontrivial matrix element
\be
\langle \Psi_{+3/2}^\mathrm{pure} | \mathcal{H} |  \Psi_{-1/2}^\mathrm{pure} \rangle = \alpha = \langle \Psi_{+1/2}^\mathrm{pure} | \mathcal{H} |  \Psi_{-3/2}^\mathrm{pure} \rangle.
\ee
Subtracting the constant $-5/4$, the four-by-four matrix of the Hamiltonian $\mathcal{H}$ is
\be
\label{eq:H44_simple_App}
\begin{tabular}{c|c}
\begin{tabular}{cccc}
$\scriptstyle{+3/2}$ \\$\scriptstyle{+1/2}$\\$\scriptstyle{-1/2}$\\$\scriptstyle{-3/2}$
\end{tabular} 
&
$\left(
\begin{tabular}{cccc}
$-1$ & $0$ & $\alpha$ & $0$\\
$0$ & $1$ & $0$ & $\alpha$\\
$\alpha^*$ & $0$ & $1$ & $0$\\
$0$ & $\alpha^*$ & $0$ & $-1$
\end{tabular} 
\right).$
\end{tabular}
\ee
It is composed of two isolated two-by-two blocks, mapping to each other by the TRS,
\be
\begin{tabular}{c|c}
\begin{tabular}{cc}
$\scriptstyle{+3/2}$ \\$\scriptstyle{-1/2}$
\end{tabular} 
&
$\left(
\begin{tabular}{cc}
$-1$ & $\alpha$\\
$\alpha^*$ & $1$
\end{tabular} 
\right)$
\end{tabular}
\overset{TRS}{\longleftrightarrow}
\begin{tabular}{c|c}
\begin{tabular}{cc}
$\scriptstyle{-3/2}$\\$\scriptstyle{+1/2}$
\end{tabular}
&
$\left(
\begin{tabular}{cc}
$-1$ & $\alpha^*$\\
$\alpha$ & $1$
\end{tabular} 
\right).$
\end{tabular}
\ee
Correspondingly, the Hamiltonian eigenstates come in pairs mapping to each other by the TRS. The pair with the higher eigenvalue, 
\be
\label{eq:eigen epsilon}
\epsilon = \sqrt{1+|\alpha|^2},
\ee
is 
\begin{subequations}
\be \begin{split}
&| \Psi_{+1/2} \rangle = (\sin\vartheta/2, 0, e^{i\varphi} \cos\vartheta/2,0)^T, %
\\&
| \Psi_{-1/2} \rangle = (0, e^{-i\varphi} \cos\vartheta/2, 0,\sin\vartheta/2)^T,
\end{split}
\ee
and the pair with the lower eigenvalue, $-\epsilon$, is
\label{eq:second step}
\be \begin{split}
\label{eq:second step HH}
&|\Psi_{+3/2} \rangle = (\cos\vartheta/2, 0, -e^{i\varphi} \sin\vartheta/2,0)^T, %
\\&
| \Psi_{-3/2} \rangle = (0, -e^{-i\varphi} \sin\vartheta/2, 0,\cos\vartheta/2)^T.
\end{split}
\ee
In these equations, 
\begin{align}
\label{eq:theta cos}
\cos \vartheta &=\frac{1}{\sqrt{1+|\alpha|^2}},\\
\label{eq:theta sin}
\sin \vartheta &= \frac{|\alpha|}{\sqrt{1+|\alpha|^2}},\\
\label{eq:eigen phi}
\tan \varphi&=-\frac{\mathrm{Im}(\alpha)}{\mathrm{Re}(\alpha)},
\end{align}
\end{subequations}
so that $\vartheta \in [0,\pi/2]$ and $\varphi \in[0,2\pi]$ for a general complex number $\alpha$. The definition in Eq.~\eqref{eq:eigen phi} is consistent with the definition in Eq.~\eqref{eq:varphi}.

\subsection{Orthogonal transformations of matrix $C$}

Since the matrix $\dotAxis{C}$ elements are the coefficients of a rank-two tensor [see Eq.~\eqref{eq:HC correspondence}], they transform under coordinate rotations according to, essentially, Eq.~\eqref{eq:coordinateRotation}:
\be
c_{\dotAxis{i}\dotAxis{j}} = R_{ik} R_{jl} c_{\deflectedAxis{k}\deflectedAxis{l}}. 
\ee
In the matrix notation, we can write
\be
\dotAxis{C} = R \deflectedAxis{C} R^T.
\ee 
The orthogonal transformations of $\dotAxis{C}$, and among them, the process of diagonalization of $\dotAxis{C}$, can therefore be interpreted as changing the Cartesian coordinate frame for the spin operators $\mathbf{J}$. The matrix $\dotAxis{C}$ can be diagonalized by an orthogonal transformation $R[\phi, \theta_+,\phi^\prime]$ parametrized by three Euler angles. These three free parameters are needed to zero the three off-diagonal entries, $c_{\dotAxis{x}\dotAxis{z}},c_{\dotAxis{y}\dotAxis{z}},c_{\dotAxis{x}\dotAxis{y}}$. Alternatively, one can bring the matrix into an `almost-diagonal form',
\be
\label{eq:H33_simple}
\left(
\begin{tabular}{ccc}
$c_{\deflectedAxis{x}\deflectedAxis{x}}$ & $c_{\deflectedAxis{x}\deflectedAxis{y}}$ & $0$\\
$c_{\deflectedAxis{x}\deflectedAxis{y}}$ & $c_{\deflectedAxis{y}\deflectedAxis{y}}$ & $0$\\
$0$ & $0$ & $c_{\deflectedAxis{z}\deflectedAxis{z}}$
\end{tabular} 
\right),
\ee
with only two free parameters, using Euler matrix $R[\phi, \theta_+ ,0]$. 
It is easy to diagonalize $\dotAxis{C}$ in Eq.~\eqref{eq:H33_simple} by an additional rotation within the $\deflectedAxis{x}\deflectedAxis{y}$ plane. We have not considered this last rotation in the main text, since it does not change the direction of the axis $\deflectedAxis{z}$ and is immaterial for the Hamiltonian form given in Eq.~\eqref{eq:H44_simple_App}: the rotation changes only the phase of $\alpha$.

We now recall the following identity of rotations of spinors and vectors
\be
\label{eq:rotationIdentity}
\mathbf{J} \cdot R \mathbf{n} = U \mathbf{J} U^\dagger \cdot \mathbf{n}, 
\ee
where $\mathbf{n}$ is a unit vector, $\mathbf{J}$ is a three-component vector of spin-3/2 operators, $R$ is a three-by-three matrix implementing a rotation in the Cartesian space, and $U$ is the corresponding rotation in the spin space,
\begin{align}
\label{eq:RU_correspondence}
R = \exp( -i \mathbf{m} \cdot \mathbf{l} \gamma) \longleftrightarrow U = \exp( -i \mathbf{m} \cdot \mathbf{J} \gamma),
\end{align}
with $\mathbf{m}$ the rotation axis and $\gamma$ the rotation angle, and $\mathbf{l}$ a three-component vector of rotation generators in three-dimensional space (three by three matrices).
Upon a coordinate rotation $R$, the two corresponding forms of the Hamiltonian change as follows
\be
\mathbf{J}^T \cdot \underbrace{R^T\dotAxis{C} R}_{\deflectedAxis{C}} \cdot \mathbf{J} = \underbrace{U^\dagger \mathbf{J}^T \cdot \dotAxis{C} \cdot \mathbf{J} U}_{\deflectedAxis{H}},
\ee
with the corresponding operators given in Eq.~\eqref{eq:RU_correspondence}. 

We conclude that one way to diagonalize a spin-3/2 Hamiltonian with TRS is the following two-step procedure. The first step is equivalent to finding a suitable coordinate frame: One simplifies the matrix $\dotAxis{C}$, bringing it to the form in Eq.~\eqref{eq:H33_simple}. The Hamiltonian expressed in pure heavy- and light-hole states then takes the form of Eq.~\eqref{eq:H44_simple_App}. The latter (four-by-four) matrix cannot be diagonalized further 
within the class of considered transformations $U$, being the one in Eq.~\eqref{eq:RU_correspondence}. While the Hamiltonian can be diagonalized [the result given in Eq.~\eqref{eq:second step}], the unitary transformation implementing it is not a Cartesian coordinate rotation, as can be inferred from Eq.~\eqref{eq:H_s}.

\subsection{Physical meaning of the canonical coordinate frame}

Here we show that the canonical coordinate frame in which the Hamiltonian is `simple', taking the form of Eq.~\eqref{eq:H44_simple_App}, has a physical meaning. That is, it is not just an arbitrary frame suitable for some calculations, but it can be identified as a solution to well-posed physical problems. Specifically, we show that the direction $\deflectedAxis{z}$ is 1) the direction of the vector $\mathbf{n}$ in Eq.~\eqref{eq:measure1} that minimizes $m_1$ (thus, maximizes the $g$ factor), and simultaneously 2) the direction of the vector $\mathbf{n}$ in Eq.~\eqref{eq:measure2} that minimizes $m_2$ (thus, maximizes the pure heavy-hole wave-function components), and simultaneously 3) the `$z$' axis of a coordinate frame in which the Hamiltonian matrix in Eq.~\eqref{eq:H44_simple_App} has the smallest possible sum of off-diagonal elements squared. We now prove these three properties.

To prove 1), we consider
\be
m_1 =\frac{3}{2} - \max_{\mathbf{n}} \langle \Psi_{3/2} | \mathbf{n} \cdot \mathbf{J} | \Psi_{3/2}\rangle.
\ee
Since the maximum is searched over all directions, we can consider the doubly-primed coordinate system
\be
m_1 = \frac{3}{2} - \max_{\deflectedAxis{\mathbf{n}}} \langle \Psi_{3/2} | \deflectedAxis{\mathbf{n}} \cdot \deflectedAxis{\mathbf{J}} | \Psi_{3/2}\rangle.
\ee
In this coordinate frame, the exact eigenspinor is simple,
\be
| \Psi_{3/2}\rangle = \cos\vartheta/2 | \Psi_{3/2}^\mathrm{pure}\rangle - e^{i \varphi}\sin\vartheta/2 | \Psi_{-1/2}^\mathrm{pure}\rangle.
\ee
Since the operator $\mathbf{n} \cdot \mathbf{J}$ is linear in $J_\pm$, the cross terms are zero and the expression simplifies to
\be \begin{split}
m_1 &= \frac{3}{2} - \max_{\deflectedAxis{\mathbf{n}}}
 \langle \Psi_{3/2} | \cos^2(\vartheta/2) n_\deflectedAxis{z} J_\deflectedAxis{z} | \Psi_{3/2}\rangle \\ &\qquad +  \langle \Psi_{-1/2} | \sin^2(\vartheta/2) n_\deflectedAxis{z} J_\deflectedAxis{z} | \Psi_{-1/2}\rangle\\
  & = \frac{3}{2} - \max_{\deflectedAxis{\mathbf{n}}} n_\deflectedAxis{z} \left( \frac{3}{2} \cos^2(\vartheta/2) -\frac{1}{2} \sin^2(\vartheta/2)  \right).
\end{split}
\ee
Since the factor in brackets is non-negative for any $\vartheta$ within its domain, $\vartheta \in [ 0,\pi/2]$, the maximum is reached for $\deflectedAxis{\mathbf{n}} = (0,0,1)$, where
\be
m_1 =  
2 \sin^2 (\vartheta/2).
\ee
That is, the maximal $g$ factor is achieved with the magnetic field along the `$z$' axis of the canonical frame.

To prove 2), we start with the definition
\be
m_2 = 1-\frac{1}{2}\max_{U} \mathrm{Tr} \rho_{3/2} U \rho^{0}_{3/2} U^\dagger,
\ee
and exploit the canonical coordinate frame $\rho_{3/2} = V \rho_{3/2}^\mathrm{simple} V^\dagger$, where $V$ implements the transformation into the canonical coordinate frame, and $\rho_{3/2}^\mathrm{simple}$ is the projector on the subspace of eigenstates given in Eq.~\eqref{eq:second step HH}. Since the set over which the right-hand side is minimized is invariant with respect to a fixed coordinate rotation $V$, we get
\be
m_2 = 1 - \frac{1}{2}\max_{W} \mathrm{Tr} \rho_{3/2}^\mathrm{simple} W  \rho^{0}_{3/2} W^\dagger.
\ee
We now insert the definition given below Eq.~\eqref{eq:measure2}, $\rho_{3/2}^0 = \mathrm{diag}(1,0,0,1)$ and parametrize the searched-for rotation by two Euler angles, $W = \exp[ -i v (J_\deflectedAxis{x} \cos u + J_\deflectedAxis{y} \sin u) ]$, exploiting the fact that the third rotation, along the resulting $\deflectedAxis{z}$, leaves the projector $\rho^{0}_{3/2}$ invariant. By explicit evaluation and some algebraic manipulations, we got the expression
\be \begin{split}
m_2 = 1- \frac{1}{8} \max_{u,v} \Big\{ &4 + \cos \vartheta +3 \cos \vartheta \cos 2 v \\
&\quad + 2\sqrt{3}  \sin \vartheta \sin^2 v \cos (2u-\varphi) \Big\}.
\end{split}
\ee
Here, the angles $\vartheta$ and $\varphi$ appear from having expressed $\rho_{3/2}^\mathrm{simple}$ using Eq.~\eqref{eq:second step HH}, and the two angles $u, v$ are parameters over which the expression is maximized. Since the first three terms in the curly brackets add to a non-negative number, the expression will be maximal if the last term is the largest positive possible. The condition selects $2u-\varphi=0$ mod $2\pi$. The expression becomes
\be
m_2 = 1 - \frac{1}{2}\max_{v} \Big\{ 1 + \cos \vartheta - \frac{1}{2} \sin^2 v \left( 3 \cos \vartheta-  \sqrt{3} \sin \vartheta  \right) \Big\},
\ee
which simplifies to 
\be
m_2 = \frac{1-\cos \vartheta}{2} + \frac{1}{4}\min_{v} \Big\{ \sin^2 v \left( 3 \cos \vartheta-  \sqrt{3} \sin \vartheta  \right) \Big\},
\ee
The expression in the round bracket is 
positive for $\vartheta \in [0, \pi/3]$ and negative for $\vartheta \in [\pi/3, \pi/2]$. In the first case, we get the maximum at $v=0$ mod $\pi$ with value
\be
m_2 = \sin^2 (\vartheta/2), \quad \mathrm{if} \, \vartheta \leq \frac{\pi}{3}.
\ee
The second case gives $v = \pi/2$ mod $\pi$ and
\be
m_2 = \frac{1}{4}\left( 2 + \cos \vartheta - \sqrt{3} \sin\vartheta \right), \quad \mathrm{if} \, \vartheta \geq \frac{\pi}{3}.
\ee
Thus, if the mixing angle is not large, $\vartheta \leq \frac{\pi}{3}$, the measure $m_2$ is maximized for $\mathbf{n}$ in Eq.~\eqref{eq:measure2} being along the canonical frame $\deflectedAxis{z}$ axis.
 
The crossover (a qualitative change) at $\alpha =\sqrt{3}$ corresponding to $\vartheta=\pi/3$ that $m_2$ displays can be understood by looking at the matrix $C$ in the canonical frame, where this matrix is diagonal. For small mixing, the matrix is $\deflectedAxis{C}=\mathrm{diag}(A, -A, -1)$, with $A=\pm |\alpha|/\sqrt{3}$ small. At the crossover, when $|\alpha| = \sqrt{3}$, the value of $A$ reaches $\pm1$. For even larger $|\alpha|$, the component $c_{\deflectedAxis{z}\deflectedAxis{z}}$ is not anymore the dominant one. One should rename the axes, taking $\deflectedAxis{y}$ (or $\deflectedAxis{x}$, depending on the sign of $A$) as the new `$z$' axis. This axes renaming converts a $\deflectedAxis{C}$ matrix with $|\alpha| >\sqrt{3}$ to a matrix with $|\alpha| \leq \sqrt{3}$. The explicit mapping is 
\be
\label{eq:alphaMapping}
|\alpha| \to \frac{\sqrt{3}+|\alpha|}{\sqrt{3}|\alpha|-1},
\ee
which translates to $\vartheta \to 2\pi/3-\vartheta$.
Considering the availability of this mapping, the mixing parameter values larger than $\pi/3$ are not relevant. %

To prove 3), we consider the four-by-four matrix $H$ in Eq.~\eqref{eq:HC correspondence} and the expression in Eq.~\eqref{eq:measure22},
\be
m_3= \frac{1}{\Delta^{2}}\min_{\mathbf{n}} \sum_{i\neq j} |h^{\mathbf{n}}_{ij}|^2.
\ee
We aim to prove that $\mathbf{n}$ along $\deflectedAxis{z}$ minimizes $m_3$. To this end, let us take the sum of all matrix elements squared
\be
\sum_{ij} |h_{ij}|^2 = \sum_{ij} h_{ij} h_{ij}^* =\sum_{ij} (H)_{ij} (H^\dagger)_{ji} = \mathrm{Tr} \left (H H^\dagger \right).
\ee
This expression is invariant with respect to unitary transformations of $H$ (thus, also to Cartesian coordinate rotations). Indeed,
\be
\mathrm{Tr} \left (H H^\dagger \right)  =  \mathrm{Tr} \left (U H U^\dagger U H U^\dagger \right)  = \mathrm{Tr} \left (H^\prime  H^{\prime  \dagger} \right). 
\ee
Since the sum of all elements squared is a constant, the minimization of the sum of off-diagonal elements is equivalent to the maximization of the sum of the on-diagonal elements.

From the six operators constituting a basis for $H$ given in Eq.~\eqref{eq:TRS_basis}, only $ \mathbb{1}$ and $J_\someAxis{z}^2$ have matrix elements on the diagonal. The sum of the squared off-diagonal elements of $H$ then is
\be
\label{eq:D33}
\begin{split}
m_3 &= \Delta^{-2}\mathrm{Tr}(H^2) - \\&\Delta^{-2}\max_{\someAxis{\mathbf{n}}} \left\{ 2 \left( \frac{15}{4}c_{\someAxis{\mathbb{1}}} + \frac{9}{4}c_{\someAxis{z}\someAxis{z}} \right)^2 + 2 \left( \frac{15}{4} c_{\someAxis{\mathbb{1}}} + \frac{1}{4} c_{\someAxis{z}\someAxis{z}}  \right)^2 \right\},
\end{split}
\ee
where the tilde means $\someAxis{\mathbf{n}}$ is an arbitrary unit vector, not related to any of the considered coordinate frames in any special way. 

We now consider a Hamiltonian expressed in the canonical frame with a diagonal associated matrix $\deflectedAxis{C} = \mathrm{diag} (c_{\deflectedAxis{x}\deflectedAxis{x}}, c_{\deflectedAxis{y}\deflectedAxis{y}}, c_{\deflectedAxis{z}\deflectedAxis{z}})$, and examine how does the sum of the diagonal elements squared behave upon adopting a different coordinate frame. That is, we evaluate Eq.~\eqref{eq:D33} for a transformed matrix
\be
\label{eq:CtoC}
\someAxis{C} = R[u, v,u^\prime]^T \deflectedAxis{C} R[u, v,u^\prime],
\ee
for arbitrary angles $u$, $v$, $u^\prime$. Here, it is the only place in this paper where one needs to pay attention to the gauge freedom explained below Eq.~\eqref{eq:gaugeDOFcomment}. The constraint $c_{\deflectedAxis{y}\deflectedAxis{y}}+c_{\deflectedAxis{x}\deflectedAxis{x}}=0$ is not preserved by Eq.~\eqref{eq:CtoC}. We assume $c_\deflectedAxis{\mathbb{1}}=0$ and assign $c_\someAxis{\mathbb{1}} = (c_{\someAxis{x}\someAxis{x}} + c_{\someAxis{y}\someAxis{y}})/2$ as explained below Eq.~\eqref{eq:gaugeDOFcomment}.
With these definitions, we explicitly evaluate Eq.~\eqref{eq:CtoC} and after some algebraic manipulations we obtain
\begin{subequations}
\begin{align}
c_{\someAxis{\mathbb{1}}} &= \frac{1}{2} \sin^2v \left(c_{\deflectedAxis{z}\deflectedAxis{z}}-c_{\deflectedAxis{x}\deflectedAxis{x}}\cos2u\right),\\ 
c_{\someAxis{z}\someAxis{z}} &= c_{\deflectedAxis{z}\deflectedAxis{z}} \cos^2v - \frac{1}{2}\sin^2v \left(c_{\deflectedAxis{z}\deflectedAxis{z}} -3 c_{\deflectedAxis{x}\deflectedAxis{x}} \cos2u\right),
\end{align}
\end{subequations}
and finally
\be
\label{eq:m3stepX}
m_3 = \Delta^{-2} \mathrm{Tr}(H^2) -  \max_{v,u} \frac{c_{\deflectedAxis{z}\deflectedAxis{z}}^2}{\Delta^{2}} \left( \frac{41}{4} - 4y + y^2 \right),
\ee
where
\be
y = \sin^2v \left( 3 + \sqrt{3} \tan \vartheta \cos 2u \right).
\ee
The expression in the second bracket in Eq.~\eqref{eq:m3stepX}, a function of $y$, has a maximum at $y=0$ for $0 \leq y \leq 4$. This solution means the sum of the squared off-diagonal elements is minimal for $\sin v=0$, meaning in the canonical frame, as long as the mixing is not large, specifically, $\vartheta \leq  \pi/6$. In this range, the value of $m_3$ is 
\be
m_3 = \tan^2\vartheta,
\ee
what we obtained by evaluating the trace using $H$ in the canonical frame, Eq.~\eqref{eq:H44_simple_App} and putting $\Delta = -2 c_{\deflectedAxis{z}\deflectedAxis{z}}$, arriving at Eq.~\eqref{eq:m4Result}.
For $\vartheta >\pi/6$, the minimum is reached for $\sin v=1$, $\cos 2u =1$ and has the value
\be
m_3 = \frac{1}{4} (\sqrt{3} - \tan \vartheta)^2.
\ee
The crossover at $\vartheta=\pi/6$ that $m_3$ displays has the following meaning. At the crossover at $\vartheta=\pi/3$, discussed around Eq.~\eqref{eq:alphaMapping}, there is an alternative axes renaming. Namely, upon subtracting a constant, the Hamiltonian is proportional to $+J_x^2$ and describes a system where the pure light holes are the ground state. Irrespective of the light hole or heavy hole character, for $m_3$ the important fact is that such Hamiltonian is diagonal (that is, the axes can be chosen such that $\alpha=0$). It is this choice of axes that $m_3$ selects as the one minimizing the off-diagonal elements for $\vartheta>\pi/6$. Concerning the form of the Hamiltonian, adopting this choice means taking Eq.~\eqref{eq:H44_simple_App} with inverted signs on the diagonal, and mapping the mixing angle $\vartheta \to \pi/3-\vartheta$.

As a final comment, we mention that the canonical frame also minimizes the count of the non-zero off-diagonal elements. This minimal count is four, stemming from a single non-zero off-diagonal matrix element appearing in two copies required by the matrix hermiticity and further doubled by TRS, resulting in the form given in Eq.~\eqref{eq:H44_simple_App}. Thus, the canonical frame minimizes both the count as well as the total sum of the squares of the off-diagonal elements of $H$.

\subsection{Exact formulas}

\label{app:exactFormulas}

\begin{table}
\begin{tabular}{@{\quad}c@{\qquad}c@{\quad}c@{\quad}c@{\quad}c@{\quad}c@{\quad}}
\toprule
\addlinespace[0.3em]
&\multicolumn{2}{c}{scenario} &&\multicolumn{2}{c}{strain} \\
\cmidrule{2-3}  \cmidrule{5-6} 
& [llh] & [0lh] && $\epsilon_{\dotAxis{x}\dotAxis{y}}$ & $\epsilon_{\dotAxis{x}\dotAxis{z}}$ \\
\addlinespace[0.3em]
\midrule
\addlinespace[0.3em]
$\phi$ & $\pi/4$ &$ \pi/2$ &&$\pi/4$ &$0$ \\ 
\addlinespace[0.3em]
$c_{\someAxis{x}\someAxis{x}}$ & $c_{\dotAxis{y}\dotAxis{y}}$ &$ c_{\dotAxis{y}\dotAxis{y}}$ &&$c_{\dotAxis{x}\dotAxis{y}}/2$ &$0$ \\ 
\addlinespace[0.3em]
$c_{\someAxis{y}\someAxis{y}}$ & $0$ &$0$ &&$-c_{\dotAxis{x}\dotAxis{y}}/2$ &$0$\\ 
\addlinespace[0.3em]
$c_{\someAxis{z}\someAxis{z}}$ & $c_{\dotAxis{z}\dotAxis{z}}$ &$ c_{\dotAxis{z}\dotAxis{z}}$ &&$c_{\dotAxis{z}\dotAxis{z}}$ &$c_{\dotAxis{z}\dotAxis{z}}$\\ 
\addlinespace[0.3em]
$c_{\someAxis{x}\someAxis{z}}$ & $c_{\dotAxis{y}\dotAxis{z}}$ &$c_{\dotAxis{y}\dotAxis{z}}$ &&$0$ &$c_{\dotAxis{x}\dotAxis{z}}$\\
\addlinespace[0.3em]
\bottomrule
\end{tabular}
\caption{
\label{tab:exactInfo}
The information needed to obtain the matrix $\someAxis{C}$ in Eq.~\eqref{eq:matrixCT} for the four cases (given in the header line) considered in the main text. For `scenario [llh]', the matrix $C$ has  non-zero entries $c_{\dotAxis{z}\dotAxis{z}}$, $c_{\dotAxis{x}\dotAxis{z}}=c_{\dotAxis{y}\dotAxis{z}}$, $c_{\dotAxis{x}\dotAxis{y}}$ as given in Eq.~\eqref{eq:H0-firstScenario}. For `scenario [0lh]', the matrix $C$ has  non-zero entries $c_{\dotAxis{z}\dotAxis{z}}$, $c_{\dotAxis{y}\dotAxis{z}}$, $c_{\dotAxis{y}\dotAxis{y}}$, as given in Eq.~\eqref{eq:H0-secondScenario}. For `strain $\epsilon_{xy}$', the matrix $C$ has  nonzero entries given in Eq.~\eqref{eq:BirPikus2} putting $\epsilon_{yz}=0=\epsilon_{zx}$, namely $ c_{\dotAxis{z}\dotAxis{z}} = -\frac{2}{3} D_u \left(\epsilon_{zz} - \epsilon_{||}\right)$, and $c_{\dotAxis{x}\dotAxis{y}} = -\frac{2}{3} D_u^\prime \epsilon_{xy}$. For `strain $\epsilon_{xz}$', the matrix $C$ has  non-zero entries given in Eq.~\eqref{eq:BirPikus2} putting $\epsilon_{xy}=0=\epsilon_{yz}$, namely $c_{\dotAxis{z}\dotAxis{z}} = -\frac{2}{3} D_u \left(\epsilon_{zz} - \epsilon_{||}\right)$, $c_{\dotAxis{x}\dotAxis{z}} = -\frac{2}{3} D_u^\prime \epsilon_{xz}$.
}
\end{table}

Here we give exact formulas for the deflection angle $\theta_+$ and the measure $\mSOI$ for the scenarios considered in the main text. We first give the solution to a standard form of the Hamiltonian, and then list transformations that convert each scenario to the standard form.

By the standard form, we mean that the matrix parameterizing the Hamiltonian has been converted to the following form 
\be
\label{eq:matrixCT}
\someAxis{C} = \left(
\begin{tabular}{ccc}
$c_{\someAxis{x}\someAxis{x}}$ & $0$ & $c_{\someAxis{x}\someAxis{z}}/2$\\
$0$ & $c_{\someAxis{y}\someAxis{y}}$ & $0$\\
$c_{\someAxis{x}\someAxis{z}}/2$ & $0$ & $c_{\someAxis{z}\someAxis{z}}$
\end{tabular} 
\right).
\ee
It can be brought to a diagonal form $\deflectedAxis{C}=\mathrm{diag} (c_{\deflectedAxis{x}\deflectedAxis{x}}, c_{\deflectedAxis{y}\deflectedAxis{y}}, c_{\deflectedAxis{z}\deflectedAxis{z}})$ with
\begin{subequations}
\label{eq:matrixD}
\begin{align}
c_{\deflectedAxis{x}\deflectedAxis{x}}&= \frac{1}{2}\left(c_{\someAxis{x}\someAxis{x}}+c_{\someAxis{z}\someAxis{z}}\right) + \frac{1}{2}\sqrt{ (c_{\someAxis{x}\someAxis{x}}-c_{\someAxis{z}\someAxis{z}})^2 +c_{\someAxis{x}\someAxis{z}}^2 },\\ 
c_{\deflectedAxis{y}\deflectedAxis{y}} &= c_{\someAxis{y}\someAxis{y}},\\
c_{\deflectedAxis{z}\deflectedAxis{z}} &= \frac{1}{2}\left(c_{\someAxis{x}\someAxis{x}}+c_{\someAxis{z}\someAxis{z}}\right) - \frac{1}{2}\sqrt{ (c_{\someAxis{x}\someAxis{x}}-c_{\someAxis{z}\someAxis{z}})^2 +c_{\someAxis{x}\someAxis{z}}^2 }.
\end{align}
\end{subequations}
by
\be
\deflectedAxis{C}=R[0,\theta_+,0]^T \tilde{C} R[0,\theta_+,0],
\ee
with $R$ the Euler rotation matrix and the angle 
\be
\label{eq:thetaPlusExact}
\theta_+ = \frac{1}{2} \arctan \left(c_{\someAxis{x}\someAxis{x}}-c_{\someAxis{y}\someAxis{y}}, -c_{\someAxis{x}\someAxis{z}}\right),
\ee
The heavy-hole--light-hole splitting defined by $\deflectedAxis{C}$ is
\be
\label{eq:exact Delta}
\Delta = \frac{3}{2} \sqrt{ (c_{\someAxis{x}\someAxis{x}}-c_{\someAxis{z}\someAxis{z}})^2 +c_{\someAxis{x}\someAxis{z}}^2 }  - \frac{c_{\someAxis{z}\someAxis{z}}+c_{\someAxis{x}\someAxis{x}}}{2} +c_{\someAxis{y}\someAxis{y}}.
\ee

After presenting this generic case, we now give explicit formulas for transformations of each scenario into Eq.~\eqref{eq:matrixCT}. In all these specific scenarios, the transformation from the $\dotAxis{x}\dotAxis{y}\dotAxis{z}$ coordinate frame to the above considered $\someAxis{x}\someAxis{y}\someAxis{z}$ frame is implemented by 
\be
\label{eq:CC}
\someAxis{C} = R[\phi,0,0]^T \dotAxis{C} R[\phi,0,0].
\ee
The angle $\phi$ as well as the matrix elements of $C$ are given in Tab.~\ref{tab:exactInfo}.

\section{Derivation of Tab.~\ref{tab:projectionRules}}

Expressing the mixing terms through operators $J_\pm$ defined in Footnote \ref{fnt:Jpm}, it is straightforward to derive the following table
\be
\begin{tabular}{@{\quad}c@{\quad}c@{\quad}c@{\quad}c@{}}
\toprule
&&\multicolumn{2}{c}{$H_\mathrm{mix}$}\\
\cmidrule{3-4}
$H_\mathrm{r}^\mathrm{\scriptscriptstyle 3D}$&&$\{J_x, J_y\}$ & $J_x^2 - J_y^2$\\
\midrule
\addlinespace[0.3em]
$J_+$&\multirow{2}{*}{$\overset{\{.,H_\mathrm{mix}\}}{\xrightarrow{\hspace{1cm}}}$}& $2J_+^3/i$ & $ 2 J_+^3$\\
\addlinespace[0.3em]
$J_-$ && $-2J_-^3/i$ & $ 2 J_-^3$\\
\bottomrule
\end{tabular}.
\ee
The body of the table gives the anticommutator $\{H_\mathrm{r}^\mathrm{\scriptscriptstyle 3D}, H_\mathrm{mix} \}$, dropping the terms that are zero upon projecting them to the heavy-hole subspace by $P_\mathrm{hh} \cdot P_\mathrm{hh}$, for two alternatives for $H_\mathrm{mix}$ and two alternatives for $H_\mathrm{r}^\mathrm{\scriptscriptstyle 3D}$ as given in the table row and column headers. Evaluating a single matrix element $\langle \Psi_{3/2}^\mathrm{pure} |J_+^3 |\Psi_{-3/2}^\mathrm{pure} \rangle=3/\sqrt{2}$, the remaining matrix elements being zero, we get 
\be
J_\pm^3 \toHH P_\mathrm{hh}J_\pm^3 P_\mathrm{hh} = 3 s_\pm.
\ee
With this, the table for projections becomes
\be
\begin{tabular}{@{\quad}c@{\quad}c@{\quad}c@{\quad}c@{}}
\toprule
&&\multicolumn{2}{c}{$H_\mathrm{mix}$}\\
\cmidrule{3-4}
$H_\mathrm{r}$&&$\{J_x, J_y\}$ & $J_x^2 - J_y^2$\\
\midrule
\addlinespace[0.3em]
$J_+$&\multirow{2}{*}{$\overset{P_\mathrm{hh}\{.,H_\mathrm{mix}\}P_\mathrm{hh}}{\xrightarrow{\hspace{1cm}}}$}& $6s_+/i$ & $ 6 s_+$\\
\addlinespace[0.3em]
$J_-$ && $-6s_-/i$ & $ 6 s_-$\\
\bottomrule
\end{tabular}.
\ee
To get Tab.~\ref{tab:projectionRules} in the main text, each entry should be multiplied by $-1/\Delta$ and the rows should be linearly combined to switch back to the Cartesian-index operators $J_x$ and $J_y$. 

\section{Derivations of the SOI  projections}

\label{app:derivation}

Here we show how to derive results such as Eqs.~\eqref{eq:projectedRashba0}, \eqref{eq:projectedRashba1}, and \eqref{eq:projectedRashba2} using the projection rules given in Tab.~\ref{tab:projectionRules}. The rules (for going from the bulk spin operators $\mathbf{J}$ to heavy-hole pseudospin operators $\mathbf{s}$) are simple in the canonical coordinate frame $\deflectedAxis{x}\deflectedAxis{y}\deflectedAxis{z}$. However, in going from the bulk to the two-dimensional hole gas, we are implicitly applying further `rules' for the momentum 
operators, namely $k_\dotAxis{z} \toTwoD 0$. One can also interpret the fact that the electric field is along the \growthDirection{} as yet another rule, $E_\dotAxis{x} \toTwoD 0$ and $E_\dotAxis{y} \toTwoD 0$. Therefore, the projection for the spin operators and for the momentum and electric field vectors are simplest in different coordinate frames. The relative rotation between these two frames will complicate the resulting expressions, in principle. However, here we show that the resulting differences are of order $O(\theta_+^2) = O(\gamma_d^2)$ and can be thus neglected within the precision that we work with, set by the perturbation order included in Eq.~\eqref{eq:effectiveHamiltonian}.

To show it, we first note that the relative rotation between the singly- and doubly-primed coordinates is given by Eq.~\eqref{eq:rotationMatrixR} with $\theta \to \thetaPlus$ and the angles $\phi^\prime=-\phi$ as specified for the two scenarios. This simplification is due to the choice $\phi^\prime=-\phi$. The elements of matrix $R$ then give overlaps between axes unit vectors in the two coordinate frames. We note that $\deflectedAxis{\mathbf{x}}\cdot \dotAxis{\mathbf{x}} = 1 + O(\thetaPlus^2)$, the same for the alignment of the $y$ axes, and $\deflectedAxis{\mathbf{x}}\cdot \dotAxis{\mathbf{y}} = O(\thetaPlus^2)$. Therefore, to precision $O(\thetaPlus^2)$, the axes $x$ and $y$ are aligned in both coordinate systems.

We now turn to Eq.~\eqref{eq:projectedRashba0}. Starting with the bulk Rashba SOI, Eq.~\eqref{eq:bulkRashba}, stripped of the overall constant, we apply the leading-order (Tab.~\ref{tab:projectionRules} for $H_\mathrm{mix}=\mathrm{none}$) projection rule for the spin operators in the doubly primed coordinate system:
\be
\label{eq:projectedRashba0Step0}
	\mathbf{J} \cdot (\mathbf{k} \times \mathbf{E}) \toHH s_\deflectedAxis{z} \deflectedAxis{\mathbf{z}} \cdot (\mathbf{k} \times \mathbf{E}).
\ee
Next, we use the cyclic property of the cross product
\be
\label{eq:projectedRashba0Step2}
	s_\deflectedAxis{z} \mathbf{k}  \cdot (  \mathbf{E} \times \deflectedAxis{\mathbf{z}}),
\ee
and use the `rule' for the electric field vector to get
\be
\label{eq:projectedRashba0Step3}
	E_\dotAxis{z} s_\deflectedAxis{z} \mathbf{k}  \cdot (  \dotAxis{\mathbf{z}} \times \deflectedAxis{\mathbf{z}}).
\ee
As the angle between the $z$ axes in the two coordinate systems is $\thetaPlus$, we arrive at Eq.~\eqref{eq:projectedRashba0}.

We now retrace these steps for the first-order corrections. Let us consider the mixing term $J_\deflectedAxis{x}^2-J_\deflectedAxis{y}^2$, corresponding to the last column of Tab.~\ref{tab:projectionRules} and omit the numerical factor $-3(c_{\deflectedAxis{x}\deflectedAxis{x}}-c_{\deflectedAxis{y}\deflectedAxis{y}})/\Delta$, so that the projection rule is $J_\deflectedAxis{x} \toHH s_\deflectedAxis{x}$, $J_\deflectedAxis{y} \toHH s_\deflectedAxis{y}$, $J_\deflectedAxis{z} \toHH 0$. We have
\be
\label{eq:projectedRashba1Step0}
	\mathbf{J} \cdot (\mathbf{k} \times \mathbf{E}) \toHH (s_\deflectedAxis{x} \deflectedAxis{\mathbf{x}}+s_\deflectedAxis{y} \deflectedAxis{\mathbf{y}}) \cdot E_\dotAxis{z}(k_\dotAxis{y} \dotAxis{\mathbf{x}}-k_\dotAxis{x} \dotAxis{\mathbf{y}}),
\ee
where to get the second bracket, we used that $\mathbf{E}=E_\dotAxis{z} \dotAxis{\mathbf{z}}$.
Neglecting the $O(\thetaPlus^2)$ terms in the dot products of the axes vectors of the two coordinate frames, we get
\be
\label{eq:projectedRashba0Step4}
		E_\dotAxis{z}( s_\deflectedAxis{x} k_\dotAxis{y}  -s_\deflectedAxis{y} k_\dotAxis{x}),
\ee
which is Eq.~\eqref{eq:projectedRashba1}. With the other mixing term, the result in Eq.~\eqref{eq:projectedRashba2} follows analogously.

\section{Time-dependent effective \growthDirection{}}

\label{app:Heff}

Here we elucidate the effective heavy-hole Hamiltonian describing the system when the angle $\theta$ is time dependent. We remind that the angle parametrizes the effective confinement direction, that is, the normal of a surface against which the confinement presses the hole. Apart from the heterostructure, the gates and environmental electric noise also contribute to the confinement potential. Considering the time-dependent value of $\theta$ is then reasonable, though its changes will typically be small.

Before generalizing to a time-dependent case, we summarize the derivation presented in the main text for constant $\theta$. We start with a band-center Hamiltonian,
\begin{equation}
\label{eq:HeffStep1}
H=H_0(\mathbf{J},\mathbf{J}) + H_\mathrm{rot}(\mathbf{J},\mathbf{J}) + H_\mathrm{mix}(\mathbf{J},\mathbf{J}) + H^\mathrm{\scriptscriptstyle 3D}(\mathbf{J} \cdot \mathbf{A}). 
\end{equation}
 With the last term, we include a generic bulk Hamiltonian with an unspecified vector $\mathbf{A}$, covering both the SOI and Zeeman terms considered in the main text. The arguments in brackets denote that the first three terms are quadratic, and the last term is linear in the spin operators $\mathbf {J}$. The terms are defined in Eqs.~\eqref{eq:H0-delfectedGrowthDirection} and classified into parts in Eq.~\eqref{eq:termsTypes}. We have also used $H_0(\mathbf{J},\mathbf{J}) = -(\Delta/2) (\mathbf{J} \cdot \mathbf{\dotAxis{z}})^2$. 
 
We write the quadratic terms as three-by-three matrices $C$,
\begin{equation}
\label{eq:HeffStep2}
H = \dotAxis{\mathbf{J}} \cdot \left(\dotAxis{C_0} + \dotAxis{C_\mathrm{rot}} + \dotAxis{C_\mathrm{mix}} \right) \dotAxis{\mathbf{J}} + H^\mathrm{\scriptscriptstyle 3D}(\dotAxis{\mathbf{J}} \cdot \dotAxis{\mathbf{A}}),
\end{equation} 
and note that their elements, if taken from Eq.~\eqref{eq:H0-delfectedGrowthDirection}, correspond to the coordinate system $\dotAxis{x}\dotAxis{y}\dotAxis{z}$. We have denoted this using a prime on all quantities.
Our central point in the main text was that upon finding a suitable direction parametrized by $\deflectedAxis{z}$, the bilinear terms can be written with the form given in the previous equation with $\deflectedAxis{C_\mathrm{rot}}=0$,
\begin{equation}
\label{eq:HeffStep3}
H = \deflectedAxis{\mathbf{J}} \cdot \left( \deflectedAxis{C_0} + \deflectedAxis{C_\mathrm{mix}} \right) \deflectedAxis{\mathbf{J}} + H^\mathrm{\scriptscriptstyle 3D}(\dotAxis{\mathbf{J}} \cdot \dotAxis{\mathbf{A}}). 
\end{equation}
Although the elements of $C$ matrices might have changed values, their form (which elements are zero) is fixed by the definition in Eq.~\eqref{eq:termsTypes}.\footnote{The simplified formulas in Eq.~\eqref{eq:thetaPlus} are derived neglecting the $O(\thetaPlus^2)$ difference between the values of the entries in $\deflectedAxis{C_\mathrm{mix}}$ and $\dotAxis{C_\mathrm{mix}}$.} Particularly, $\deflectedAxis{C_0}$ is still a matrix with a single nonzero element, corresponding to a projector $|\deflectedAxis{z}\rangle \langle \deflectedAxis{z}|$. The transformation from Eq.~\eqref{eq:HeffStep2} to \eqref{eq:HeffStep3} is a change of the basis of the three-dimensional space, that is, a rotation. We make this rotation explicit, 
\begin{equation}
\label{eq:HeffStep4}
H = \dotAxis{\mathbf{J}} \cdot R_{\thetaPlus} \left( \deflectedAxis{C_0} + \deflectedAxis{C_\mathrm{mix}} \right) R_{\thetaPlus}^T \dotAxis{\mathbf{J}} + H^\mathrm{\scriptscriptstyle 3D}(\dotAxis{\mathbf{J}} \cdot \dotAxis{\mathbf{A}}), 
\end{equation}
with $R_{\thetaPlus}^T$ being the rotation taking the axis $\dotAxis{z}$ to $\deflectedAxis{z}$, that is $R_{\thetaPlus} = R[\phi,\thetaPlus,-\phi]$ given in Eq.~\eqref{eq:coordinateRotation}. 

We now trade the three-component vector rotation for a corresponding spinor rotation $U_{\thetaPlus}$, using the identity in Eq.~\eqref{eq:rotationIdentity},
\begin{equation}
\label{eq:HeffStep5}
H = U_{\thetaPlus} \left[  \dotAxis{\mathbf{J}} \cdot \left( \deflectedAxis{C_0} + \deflectedAxis{C_\mathrm{mix}} \right) \dotAxis{\mathbf{J}} \right] U_{\thetaPlus}^\dagger + H^\mathrm{\scriptscriptstyle 3D}(\dotAxis{\mathbf{J}} \cdot \dotAxis{\mathbf{A}}). 
\end{equation}
Next, we move the last term inside the unitaries, 
\begin{equation}
\label{eq:HeffStep6}
 H = U_{\thetaPlus} \left[ \dotAxis{\mathbf{J}} \cdot \left( \deflectedAxis{C_0} + \deflectedAxis{C_\mathrm{mix}} \right) \dotAxis{\mathbf{J}}  + 
 H^\mathrm{\scriptscriptstyle 3D}(\dotAxis{\mathbf{J}} \cdot \deflectedAxis{\mathbf{A}})
  \right] U_{\thetaPlus}^\dagger,
\end{equation}
where we used $R^T_{\thetaPlus} \dotAxis{\mathbf{A}}=\deflectedAxis{\mathbf{A}}$.
This Hamiltonian makes explicit the frame that has been implicit in the derivations in the main text. As the final step towards an effective Hamiltonian valid for a time-dependent frame, we switch to a time-independent frame for the spin operator. We choose the crystallographic coordinates for such a frame, even though doubly-primed axes corresponding to an arbitrary fixed value of $\theta_0$ would be equally suitable. In analogy with the above, it results in the replacement $\thetaPlus \to \theta+\thetaPlus$ and removing primes from the spin operators in the previous equation,
\begin{equation}
\label{eq:HeffStep7}
 H = U_{\theta+\thetaPlus} \left[ \crystalAxis{\mathbf{J}} \cdot \left( \deflectedAxis{C_0} + \deflectedAxis{C_\mathrm{mix}} \right) \crystalAxis{\mathbf{J}}  + 
 H^\mathrm{\scriptscriptstyle 3D}(\crystalAxis{\mathbf{J}} \cdot \deflectedAxis{\mathbf{A}})
  \right] U_{\theta+\thetaPlus}^\dagger.
\end{equation}
We can describe the system with a time-dependent \growthDirection{} with this result. The reference frame depends on time and generates an additional Hamiltonian term
\begin{equation}
 H_U = -i \hbar U_{\theta+\thetaPlus}^\dagger \partial_t U_{\theta+\thetaPlus}.
\end{equation}
Inserting the definition of the rotation operator, we get
\begin{equation}
\label{eq:Heff}
 H_U = -\hbar \mathbf{J}\cdot \mathbf{m} \, \partial_t (\theta+\thetaPlus).
\end{equation}
This term should be added to $H^\mathrm{\scriptscriptstyle 3D}$ in Eq.~\eqref{eq:effectiveHamiltonian}. We conclude that, apart from making the vector $\deflectedAxis{\mathbf{A}}=R^T_{\theta+\thetaPlus} \crystalAxis{\mathbf{A}}$ time dependent, in general, going to the time-dependent frame transformation induces a fictitious time-dependent magnetic field $\mathbf{B}_U$, defined by
\begin{equation}
\label{eq:Beff}
2 \kappa \mu_B  \mathbf{B}_U  =  -\mathbf{m} \, \hbar \partial_t (\theta+\thetaPlus),
\end{equation}
where $\mathbf{m}=\mathbf{\dotAxis{z}}\times\mathbf{\deflectedAxis{z}}$ is a unit vector.

\section{Cubic Zeeman}

\label{app:cubic}

\newcommand{\thetaT}{\Theta}

In the crystallographic coordinates, the cubic Zeeman term is
\be
\label{eq:Hz3Definition}
H_\mathrm{z3}^\mathrm{3D} = 2\mu_B q \mathbf{B} \cdot \{ J_x^3, J_y^3, J_z^3\}.
\ee
Unlike the linear Zeeman term, the in-plane components of the magnetic field do induce nonzero matrix elements within the pure heavy-hole spinor subspace. The projection rules analogous to those given in Tab.~\ref{tab:projectionRules} as the zeroth order in $H_\mathrm{mix}$ are
$$J_x^3 \toHH \frac{3}{2} s_x, \quad J_y^3 \toHH -\frac{3}{2}s_y, \quad J_z^3 \toHH \frac{27}{4} s_z.$$ 
However, since the expression in Eq.~\eqref{eq:Hz3Definition} is not a scalar, it does not translate easily to rotated coordinate frames. For completeness, we evaluate the matrix elements of Eq.~\eqref{eq:Hz3Definition} in the heavy-hole spinor subspace (in the lowest order), and arrive to
\be
\label{eq:Hz3Result}
H_{z3}^\mathrm{2D} =  2\mu_B q \mathbf{B} \cdot \mathbf{g} \, \mathbf{s},
\ee
with the $g$ tensor 
\begin{subequations}
\be
\label{eq:z3llh}
\mathbf{g}|_{\crystalAxis{x}\crystalAxis{y}\crystalAxis{z}} =\left( \begin{tabular}{c@{\quad}c@{\quad}c}
$g_1$ & $-g_2$ & $-g_3$\\
$g_2$ & $-g_1$ & $-g_3$\\
$-g_4$ & $g_4$ & $g_5$\\
\end{tabular} \right),
\ee
with 
\begin{align}
g_1 &= \frac{3}{8}\cos^2(\thetaT/2)\left(8 \cos\thetaT- \cos2\thetaT-3 \right),\\
g_2 &= \frac{3}{8}\sin^2(\thetaT/2)\left(8 \cos\thetaT+ \cos2\thetaT+3 \right),\\
g_3 &= \frac{3}{8\sqrt{2}}\sin\thetaT \left(15-\cos 2\thetaT\right),\\
g_4 &= \frac{3}{2\sqrt{2}} \sin^3\thetaT,\\
g_5 &= \frac{3}{8}\left(17 \cos\thetaT + \cos 3\thetaT \right),
\end{align}
\end{subequations}
for the [llh] scenario and 
\begin{subequations}
\be
\label{eq:z30lh}
\mathbf{g}|_{\crystalAxis{x}\crystalAxis{y}\crystalAxis{z}} =\left( \begin{tabular}{c@{\quad}c@{\quad}c}
$g_1$ & $0$ & $0$\\
$0$ & $-g_2$ & $-g_3$\\
$0$ & $-g_4$ & $g_5$\\
\end{tabular} \right),
\ee
with 
\begin{align}
g_1 & = \frac{3}{2},\\
g_2 &= \frac{3}{2}\cos^3\thetaT,\\
g_3 &= \frac{3}{4}\sin\thetaT\left(8 - \cos2\thetaT \right),\\
g_4 &= \frac{3}{2} \sin^3\thetaT,\\
g_5 &= \frac{3}{8}\left(17 \cos\thetaT + \cos 3\thetaT \right),
\end{align}
\end{subequations}
for the [0lh] scenario. In these equations, we have used $\thetaT{}=\theta+\thetaPlus$. Also, the components of the magnetic field $\mathbf{B}$ in Eq.~\eqref{eq:Hz3Result} are supposed to be evaluated in the crystallographic frame, whereas the operator $\mathbf{s}$ refers to the heavy-hole spinors in the canonical frame, in line with conventions of Sec.~\ref{sec:4A}.

\bibliography{bib/2023-Stano-HH-LH-Mixing.bib}

\end{document}